\documentclass[journal]{IEEEtran}

\usepackage{threeparttable}
\usepackage{url}
\usepackage{array}
\usepackage[utf8x]{inputenc}
\usepackage{eqnarray}
\usepackage{graphics}
\usepackage[nointegrals]{ wasysym } 
\usepackage{siunitx}
\usepackage{amsmath} 

\newcolumntype{C}[1]{>{\centering\arraybackslash}m{#1}}
\ifCLASSINFOpdf
   \usepackage[pdftex]{graphicx}
\fi

\begin{document}
\title{Pulse Shape Discrimination of CsI(Tl) with a Photomultiplier Tube and MPPCs}

\author{Nguyen V. H. Viet,~\IEEEmembership{Member,~IEEE,}
		M. Nomachi,~\IEEEmembership{Senior Member,~IEEE,} 
		K. Takahisa, \\
		T. Shima,
		B. T. Khai,~\IEEEmembership{Member,~IEEE,} 	
		R. Takaishi, 
		and K. Miyamoto 
\thanks{This work was supported by the Grants-in-Aid for Scientific Research (Kakenhi) No. 18K03676.}	
\thanks{Nguyen V. H. Viet, B. T. Khai, R. Takaishi, and K. Miyamoto are with the Graduate School of Science, Osaka University, Toyonaka, Osaka 560-0043, Japan (email: nvhviet@rcnp.osaka-u.ac.jp).}
\thanks{M. Nomachi, K. Takahisa, and T. Shima are with the Research Center for Nuclear Physics (RCNP), Osaka University, Ibaraki, Osaka 567-0047, Japan.}
}

\maketitle
\thispagestyle{empty}

\begin{abstract}
In this study, we evaluate and compare the pulse shape discrimination (PSD) performance of multipixel photon counters (MPPCs, also known as silicon photomultiphers - SiPMs) with that of a typical photomultiplier tube (PMT) when testing using CsI(Tl) scintillators.
We use the charge comparison method, whereby we discriminate different types of particles by the ratio of charges integrated within two time-gates (the delayed part and the entire digitized waveform).
For a satisfactory PSD performance, a setup should generate many photoelectrons (p.e.) and collect their charges efficiently.
The PMT setup generates more p.e. than the MPPC setup does. 
With the same digitizer and the same long time-gate (the entire digitized waveform), the PMT setup is also better in charge collection.
Therefore, the PMT setup demonstrates better PSD performance.
We subsequently test the MPPC setup using a new data acquisition (DAQ) system. 
Using this new DAQ, the long time-gate is extended by nearly four times the length when using the previous digitizer.
With this longer time-gate, we collect more p.e. at the tail part of the pulse and almost all the charges of the total collected p.e.  
Thus, the PSD performance of the MPPC setup is improved significantly.
This study also provides an estimation of the short time-gate (the delayed part of the digitized waveform) that can give a satisfactory PSD performance without an extensive analysis to optimize this gate.
\end{abstract}


\ifCLASSOPTIONpeerreview
	\centering \bfseries EDICS Category: 3-BBND 
\fi
%
\IEEEpeerreviewmaketitle

\section{Introduction} \label{Intro}
\IEEEPARstart{M}{any} nuclear and particle physics experiments require Particle Identification (PID) to obtain the particles of interest and reduce the background radiation.
For setups using scintillators and photosensors, the most common way to perform PID is to use Pulse Shape Discrimination (PSD), which is the ability of some scintillators to emit scintillation light pulses with different shapes for different types of incident particles.

The most common photosensor used with scintillators is the photomultiplier tube (PMT).
However, the use of PMT has some limitations.
For instance, the detector environment contains He gas that can penetrate and damage the PMT. 
Alternatively, when the detector requires many photosensors in a confined size, typical PMTs with large dimensions cannot fit in the detector owing to space constraints.
Moreover, the PMT is affected by the magnetic field.
Therefore, it is unusable in some systems, such as the positron emission tomography–magnetic resonance imaging (PET-MRI) system. 
In such conditions, it requires other types of photosensors. 

A solution for the above limitations of the PMT is the use of the multipixel photon counter (MPPC), also known as silicon photomultiplier (SiPM), which is a new type of photosensor.
The PSD performance of MPPCs was demonstrated to be comparable with that of a typical PMT in the $n-\gamma$ discrimination using both organic \cite{grodzicka2018study} and inorganic \cite{mesick2015performance}, \cite{dinar2019pulse} scintillators.
In this study, we evaluate the PSD performance of MPPCs and compare it with that of a PMT in charged-particle discrimination using the CsI(Tl) scintillator.

\section{Methods and Setups} \label{Method_Setup}
\begin{figure} 
	\centering
	\includegraphics[width=0.9\linewidth]{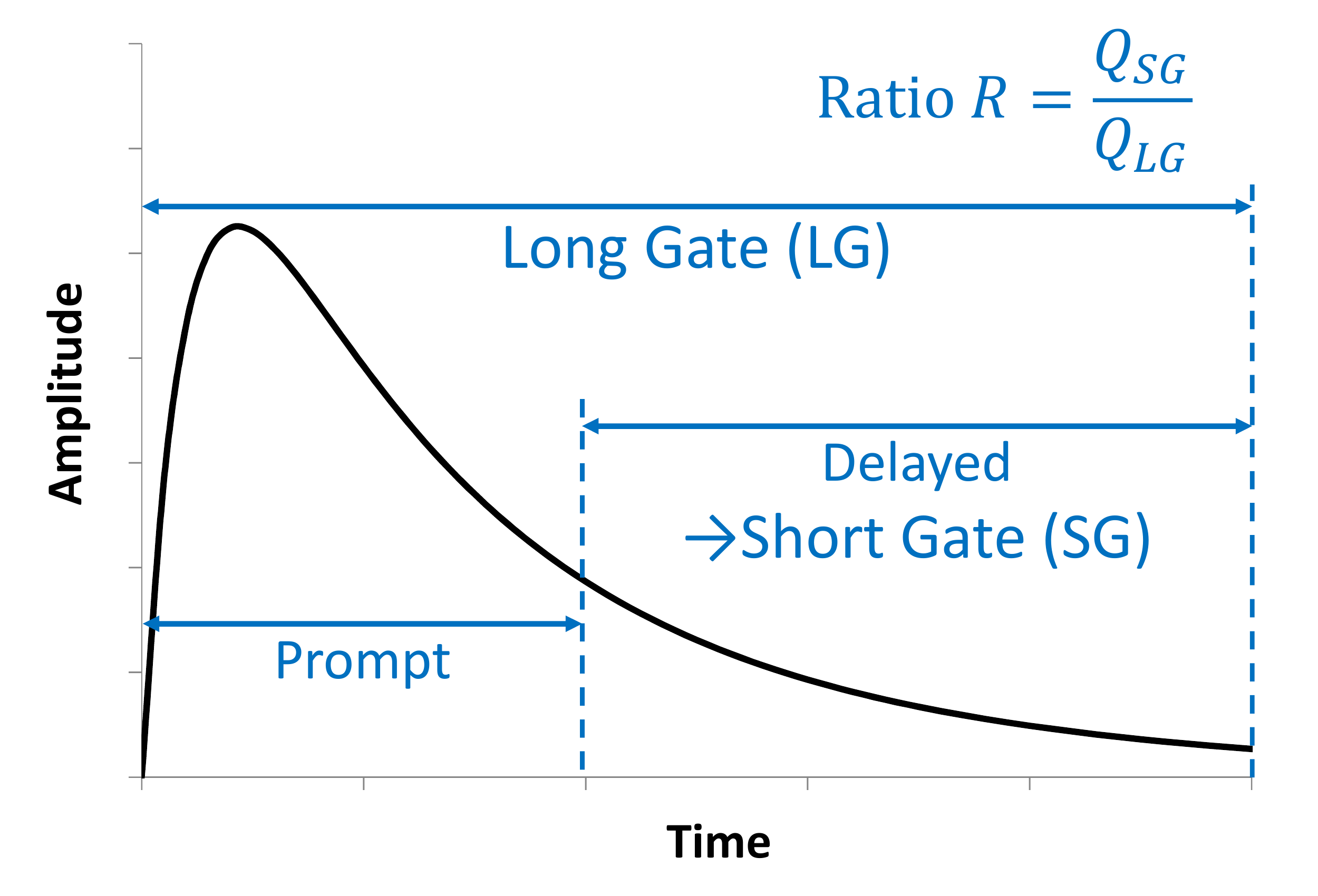}
	\caption{The charge comparison method with the definition of the Ratio R.}
	\label{fig_R_Def}
\end{figure}

To compare the PSD performances of the PMT and MPPCs, we measure the pulse of scintillation light from the PMT and MPPCs, after which we perform the charge comparison method (Fig. \ref{fig_R_Def}). 
We integrate the charges collected in the Short Gate (SG) and Long Gate (LG), the delayed part, and the entire digitized waveform, respectively. 
The Ratio R of the charges in SG and LG, $R = Q_{SG}/Q_{LG}$, identifies the type of particle.
Fig. \ref{fig_FOM_Def} shows the Ratio distributions of two types of particles with the mean value of the Ratio ($R_1$ or $R_2$) for each type and the corresponding Full Width at Half Maximum ($FWHM_1$ or $FWHM_2$).
The PSD performance for the two types of particles is evaluated using the conventional Figure of Merit (FOM), which is expressed as follows:
\begin{equation} \label{Eq_FOM_1}
FOM = \frac{\text{Peak Separation}}{FWHM_1 + FWHM_2}
= \frac{\Delta R}{\sum FWHM}
\end{equation}

{\begin{figure} 
\centering
\includegraphics[width=0.9\linewidth]{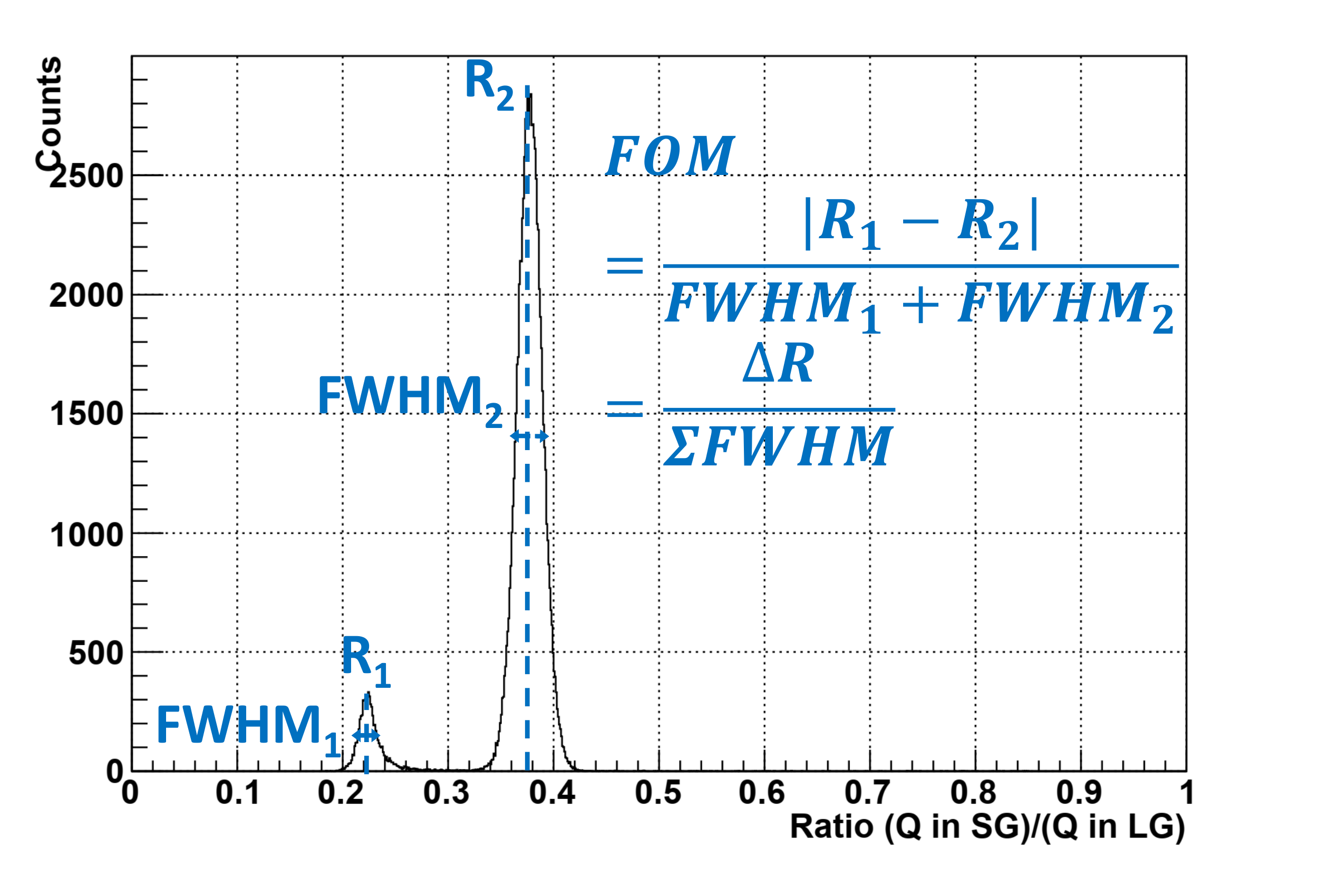}
\caption{
	Ratio distributions of two types of particles. 
	The PSD performance is evaluated using the Figure of Merit (FOM).}
\label{fig_FOM_Def}
\end{figure}}

\begin{figure}
	\centering
	\includegraphics[width=0.9\linewidth]{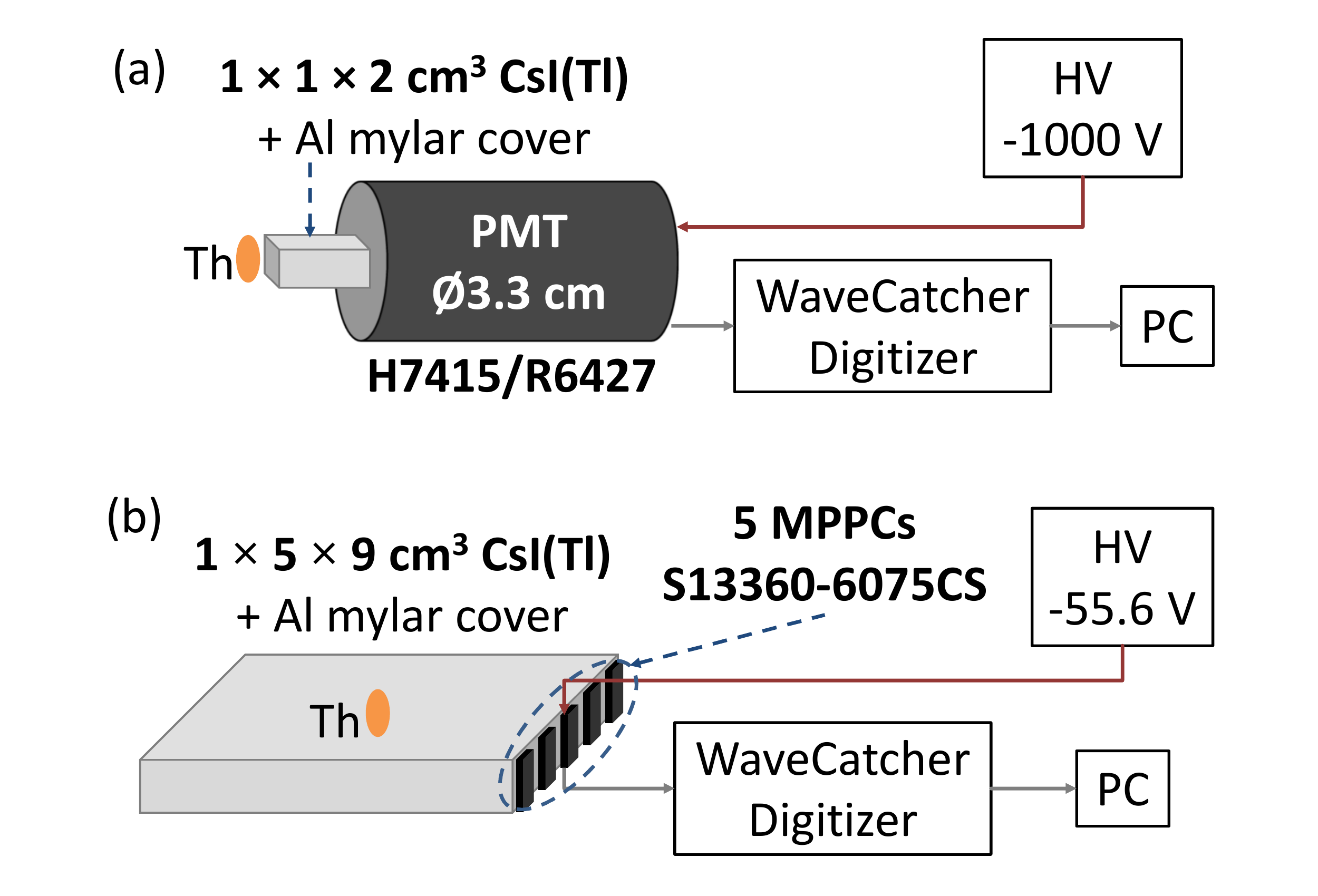}
	\caption{
		The PMT setup (top) and the MPPC setup (bottom) for PSD performance measurements. 
		For the MPPC setup, the sum signal from five MPPCs is fed into WaveCatcher.}
	\label{fig_Setup}
\end{figure}

This study mainly compares the FOMs of the PMT and MPPCs in an $\alpha-\beta$ separation.
The $\alpha-\beta$ source is natural thorium from a lantern mantle.
We use the CsI(Tl) scintillator (Leading Edge Algorithms Co., Ltd.) because of its PSD ability.
The experimental setups are shown in Fig. \ref{fig_Setup}. 
In the PMT setup, a Hamamatsu H7415/R6427 PMT \cite{pmtdatasheet} (with a diameter of \SI{33}{\milli\meter}) is coupled with a \SI[product-units = power]{10 x 10 x 20}{\milli\meter} CsI(Tl) on the \SI[product-units = power]{10 x 10}{\milli\meter} surface. 
In the MPPC setup, five Hamamatsu S13360-6075CS MPPCs \cite{mppcdatasheet} (with an effective area of \SI[product-units = power]{6 x 6}{\milli\meter} for each) are coupled with a \SI[product-units = power]{10 x 50 x 90}{\milli\meter} CsI(Tl) on the \SI[product-units = power]{10 x 50}{\milli\meter} surface. 
For the PMT, the bias voltage is \SI{-1000}{\volt}.
For each MPPC, the bias voltage is \SI{-55.6}{\volt} (\SI{3}{\volt} over breakdown, as recommended by Hamamatsu). 
In the MPPC setup, there is no temperature compensation module.
To measure the pulse of scintillation light from the PMT and the five MPPCs, we use a fast digitizer called WaveCatcher \cite{breton2014wavecatcher}, which offers a dynamic range of \SI{2.5}{\volt} over 12 bits.
The sampling depth is 1024 samples and the sampling rate ranges from 0.4 to \SI{3.2}{GS\per\second}.
Because of the long decay time of CsI(Tl) ($\sim\SI{1}{\micro\second}$), we keep the sampling rate at \SI{0.4}{GS\per\second}. 
Therefore, the range of the digitized waveform is \SI{2560}{\nano\second} on WaveCatcher.
For the MPPC setup, the signals from the five MPPCs are summed up and then fed into WaveCatcher.

We use a larger CsI(Tl) for the MPPC setup because this setup was planned for another experiment, the $\mu$ capture on $^3$He. 
This experiment represents a case whereby a typical PMT is not suitable because of the presence of He gas and the small-size chamber containing the pricey $^3$He gas.
In this $\mu$ capture experiment, we need to separate the proton ($p$) and deuteron ($d$) in the produced particles \cite{golak2014break}. 
Therefore, we also perform a $p-d$ separation measurement to confirm whether our current MPPC setup is good enough for the $\mu$ capture experiment.
This $p-d$ separation will be discussed in detail in section \ref{p-d}.

\section{PSD Estimation} \label{PSD_Estimation}
\subsection{Ratio Fluctuation}
We estimate the fluctuation of the Ratio R from the statistical distribution of the number of photoelectrons ($N_{pe}$) generated by the photosensor.
For each pulse, the charges (and $N_{pe}$) of the prompt part, delayed part, and the entire pulse are $Q_p$ ($n_p$), $Q_d$ ($n_d$), and $Q$ ($N$), respectively. 
The Ratio $R$ is represented as follows:
\begin{equation} \label{Eq_R_1}
R = \frac{Q_{SG}}{Q_{LG}}
  = \frac{n_d}{n_p+n_d}
  = \frac{n_d}{N}
\end{equation}
Because $n_p$ and $n_d$ are independent, the fluctuation of the Ratio comes from the propagation of $\sigma_{n_p}$ and $\sigma_{n_d}$, and is expressed as follows: 
\begin{equation} \label{Eq_SigmaR_Sq_1}
{\sigma_R}^2 =
\left(\frac{\partial R}{\partial n_{p}}\sigma_{n_p}\right)^2
+ \left(\frac{\partial R}{\partial n_{d}}\sigma_{n_d}\right)^2
\end{equation}
The $N_{pe}$ follows the Poisson distribution, so $\sigma_{n_p}=\sqrt{n_p}$ and $\sigma_{n_d}=\sqrt{n_d}$. 
Therefore,
\begin{equation} \label{Eq_SigmaR_Sq_2}
{\sigma_R}^2 
= \frac{1}{N^4} \left( {n_d}^2 n_p + {n_p}^2 n_d \right) 
= \frac{1}{N}R(1-R)
\end{equation}

\subsection{FOM Estimation} \label{Est_FOM}
From Eq. \ref{Eq_FOM_1} and Eq. \ref{Eq_SigmaR_Sq_2}, the estimation of the FOM from the statistical fluctuation of the $N_{pe}$ for the case of $\alpha$ and $\beta$ is expressed as follows:
\begin{equation} \label{Eq_FOM_2}
\begin{split}
FOM_{est} 
& = \frac{|R_{\beta}-R_{\alpha}|\sqrt{N}}
{2\sqrt{2\ln{2}}
	\left(
	{\sqrt{(1-R_{\beta})R_{\beta}}} +
	{\sqrt{(1-R_{\alpha})R_{\alpha}}}
	\right)
} \\
& = f(R_{\alpha},R_{\beta}) \sqrt{N}
\end{split}
\end{equation}
Therefore, the FOM is proportional to the $\sqrt{N_{pe}}$ (or $\sqrt{Q}$); and it also depends on $R_{\alpha}$ and $R_{\beta}$.
These Ratio values depend on the pulse shape, and the lengths of the SG and LG for charge integration.
For a fixed setup of the scintillator and photosensor, the FOM is maximized by optimizing the lengths of the SG and LG.
This optimization is independent of the $\sqrt{N_{pe}}$.

\section{Results and Discussion} \label{Result_Discuss}
\subsection{Charge Fluctuation} \label{Q_Fluctuation}
\begin{figure}
	\centering
	\includegraphics[width=0.9\linewidth]{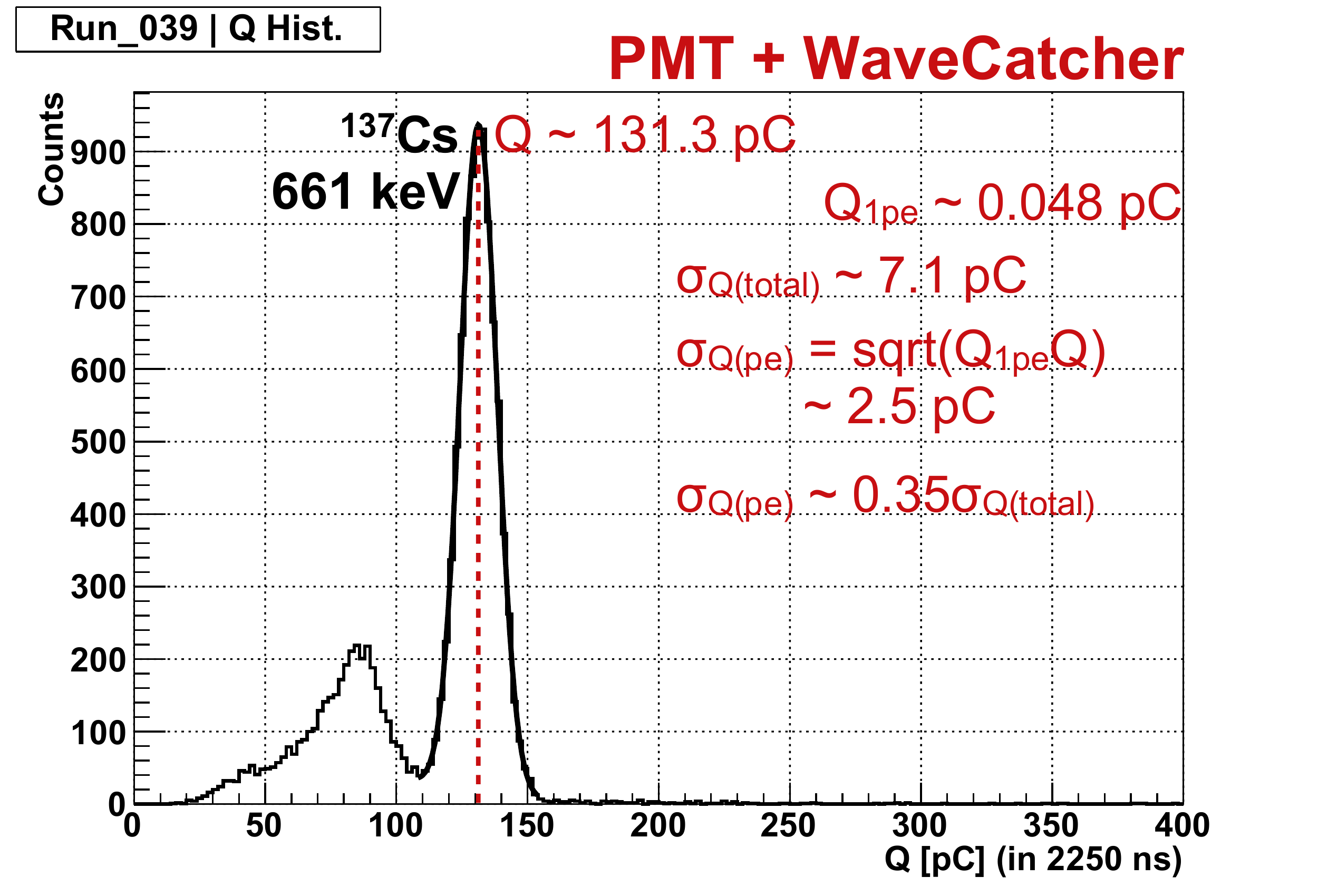}
	\includegraphics[width=0.9\linewidth]{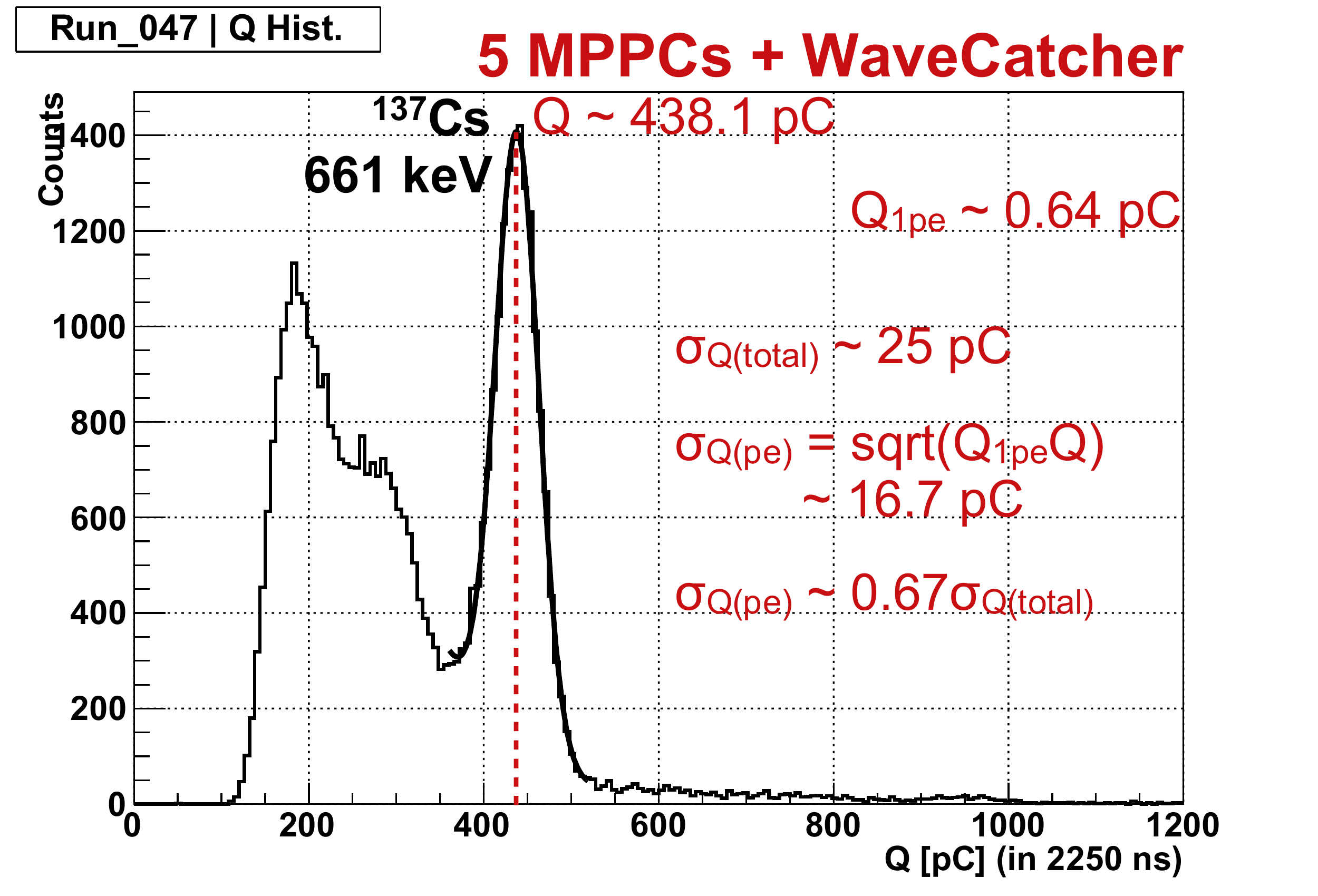}
	\caption{
		The charge histogram of $^{137}$Cs from the PMT setup (top) and the MPPC setup (bottom). 
		For the MPPC setup, the component from the $N_{pe}$ fluctuation, $\sigma_{Q(pe)}$, is the main contribution to the total charge fluctuation (the measured one), $\sigma_{Q(total)}$.
		For the PMT setup, other types of fluctuations degrade $\sigma_{Q(total)}$.
	}
	\label{fig_QHist}
\end{figure}

\begin{figure}[!h]
	\centering
	\includegraphics[width=0.9\linewidth]{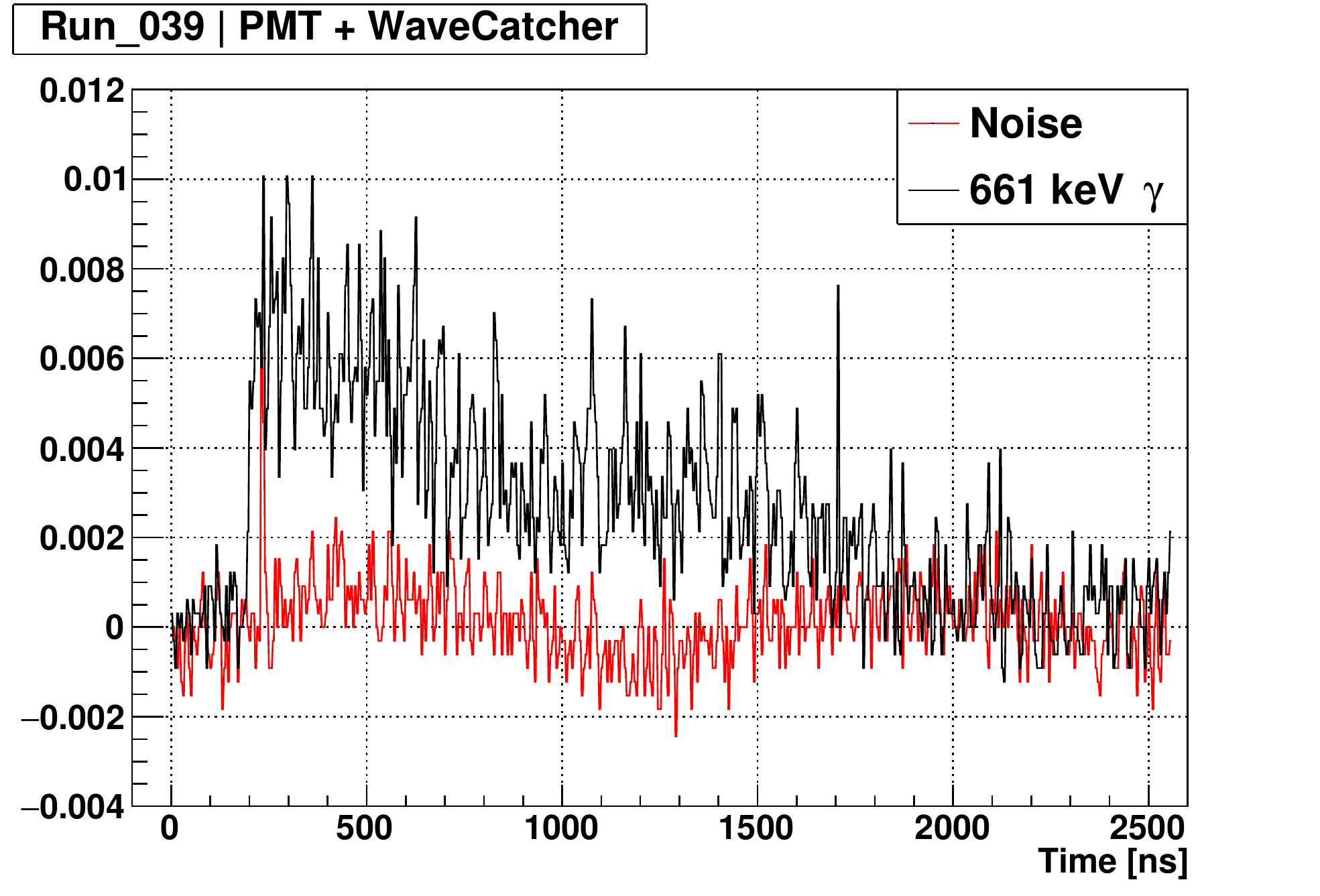}
	\caption{
		The pulse of \SI{661}{\kilo\electronvolt} $\gamma$ and the sinusoidal noise from the PMT setup.
		The period of the noise is $1300 \sim \SI{1400}{\nano\second}$.
		The amplitude of this noise is more than $10\%$ of the amplitude of the pulse of \SI{661}{\kilo\electronvolt} $\gamma$.
		This noise severely degrades the $\sigma_{Q(total)}$ of the \SI{661}{\kilo\electronvolt} peak.  
	}
	\label{fig_Wf_PMT_Gamma_Noise}
\end{figure}

First, we check the fluctuation of the charge $\sigma_{Q}$ at the peak of \SI{661}{\kilo\electronvolt} of $^{137}$Cs (Fig. \ref{fig_QHist}).
Because we cannot measure the charge of one photoelectron, $Q_{1pe}$, in our setups, it is taken from datasheets.
For the PMT, at \SI{-1000}{\volt}, $Q_{1pe}$ is approximately \SI{0.048}{\pico\coulomb}. 
The component from the $N_{pe}$ fluctuation, $\sigma_{Q(pe)}$, contributing to the total charge fluctuation (the measured one), $\sigma_{Q(total)}$, is expected as follows:
\begin{equation} \label{Eq_SigmaQ_pe_PMT}
\begin{split}
\sigma_{Q(pe)} 
& = Q_{1pe}\sqrt{N_{pe}} = \sqrt{Q_{1pe}Q} \\
& = \sqrt{0.048\times 131.3} = \SI{2.5}{\pico\coulomb} 
\end{split}
\end{equation}
where $Q = \SI{131.3}{\pico\coulomb}$ and $\sigma_{Q(total)} = \SI{7.1}{\pico\coulomb}$ represent the charge and its total fluctuation at the $^{137}$Cs peak (Fig. \ref{fig_QHist} - top).
The $\sigma_{Q(pe)}$ is less than 50\% of $\sigma_{Q(total)}$.
As shown in Fig. \ref{fig_Wf_PMT_Gamma_Noise}, there is a sinusoidal noise with a period of $1300 \sim \SI{1400}{\nano\second}$ in the PMT setup. 
The amplitude of this noise is more than $10\%$ of the amplitude of the pulse of \SI{661}{\kilo\electronvolt} $\gamma$.
It makes the resolution worse.

For the MPPC setup, $Q_{1pe}$ is approximately \SI{0.64}{\pico\coulomb}.
Through a similar calculation, we expect \SI{16.7}{\pico\coulomb} and measure \SI{25}{\pico\coulomb}.
We find no significant source of noise as the sinusoidal noise in the case of the PMT setup.

\subsection{Area-normalized Waveform} \label{Nwf_Area}
\begin{figure}
	\centering
	\includegraphics[width=0.9\linewidth]{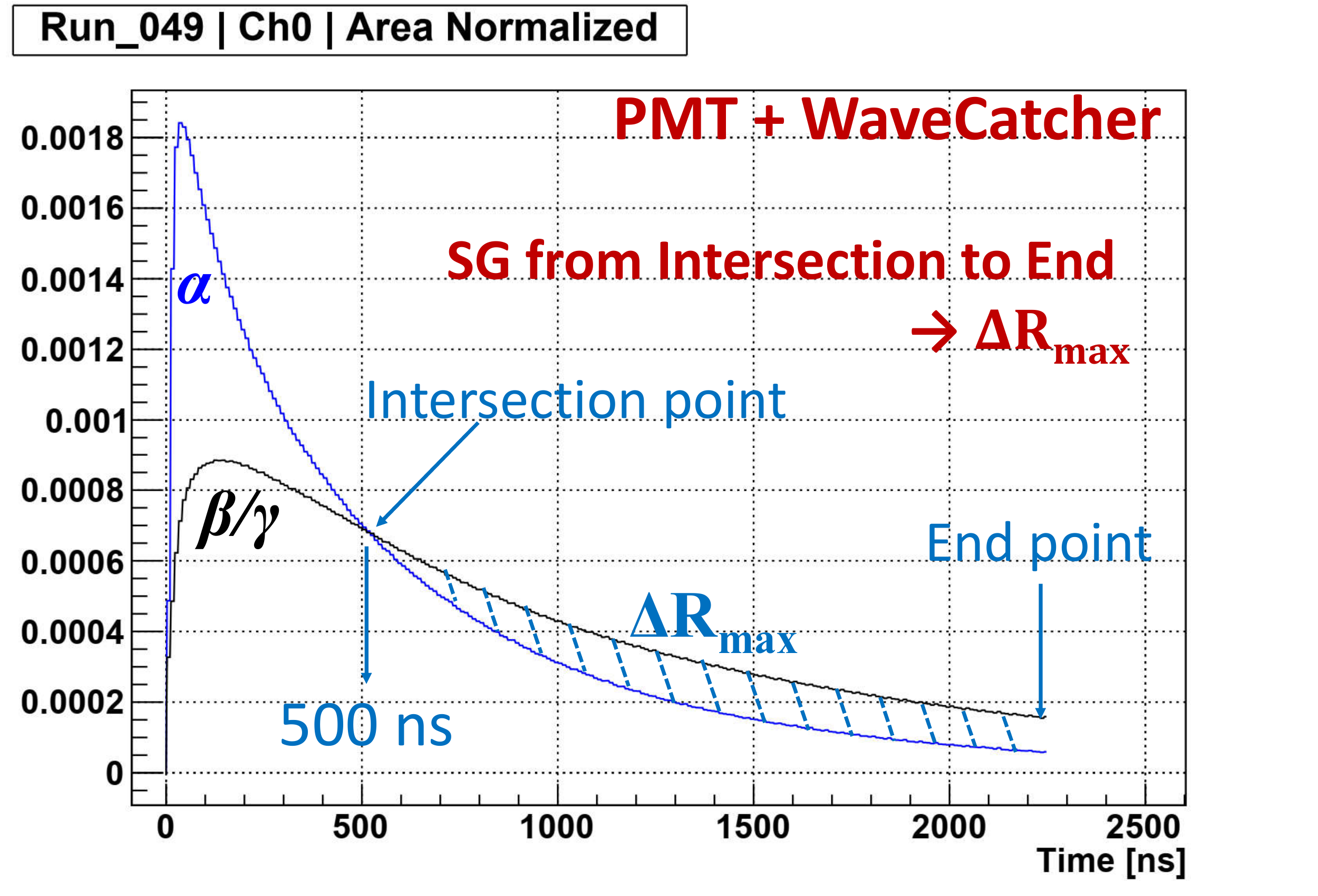}
	\includegraphics[width=0.9\linewidth]{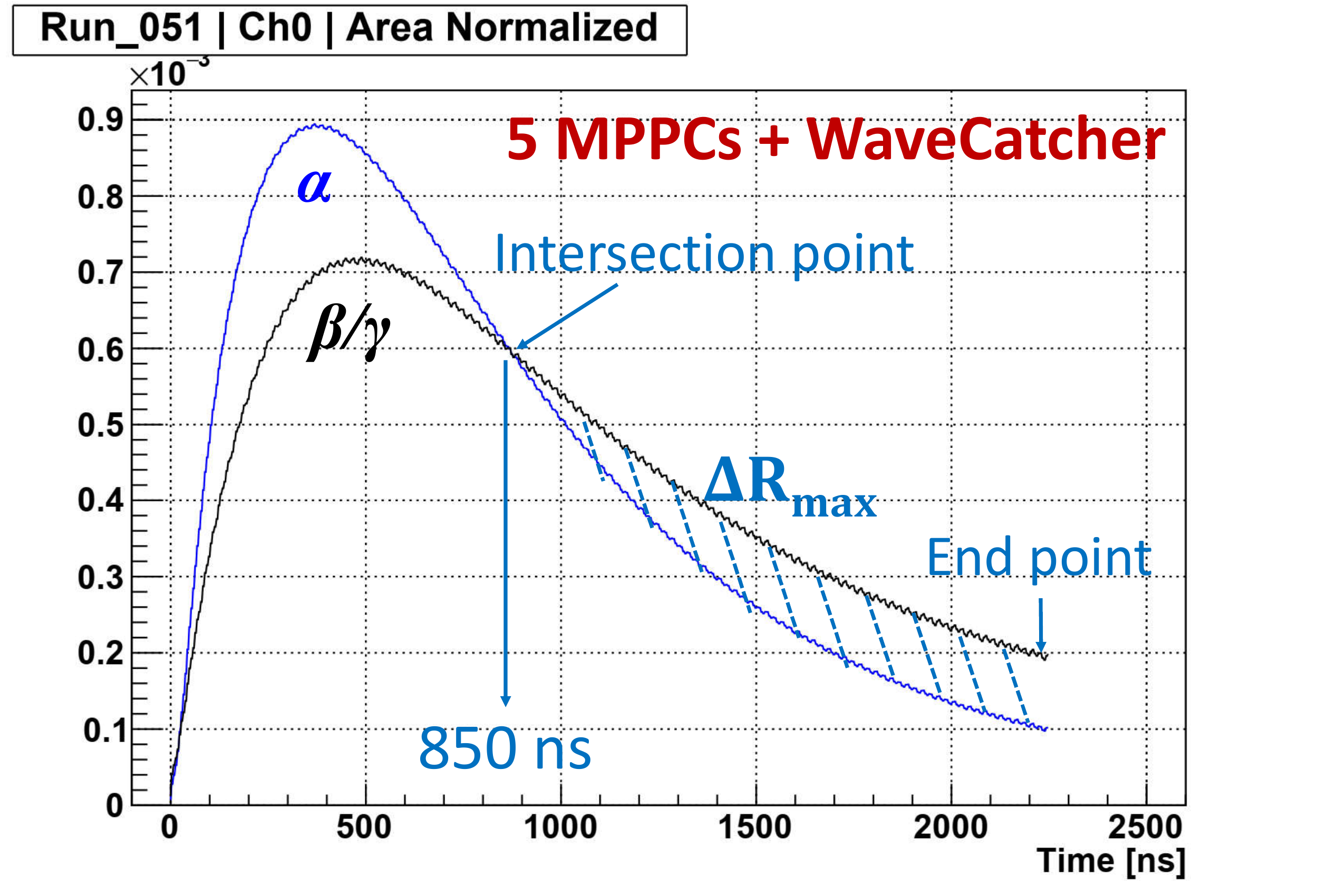}
	\caption{
		Area-normalized waveforms (normalized by the charge inside the LG) of $\alpha$ and $\beta$ from the PMT setup (top) and the MPPC setup (bottom).
		The SG from the Intersection point to the End point will give $\Delta R_{max}$, i.e., the area with dashed lines.
	}
	\label{fig_Nwf_Area_WC}
\end{figure}

We define $w_n(t) = w(t) / \sum_{LG}{w(t)}$ as the area-normalized waveform (normalized by the charge inside the LG), and thus, $R = \sum_{SG} {w_n(t)}$.
As demonstrated in Fig. \ref{fig_Nwf_Area_WC}, the area-normalized waveforms of $\alpha$ and $\beta/\gamma$ intersect at one point.
The SG from this Intersection point to the End point gives the maximum peak separation $\Delta R_{max}$, i.e., the area with dashed lines in Fig. \ref{fig_Nwf_Area_WC}. 

\begin{figure}
\centering
\includegraphics[width=0.9\linewidth]{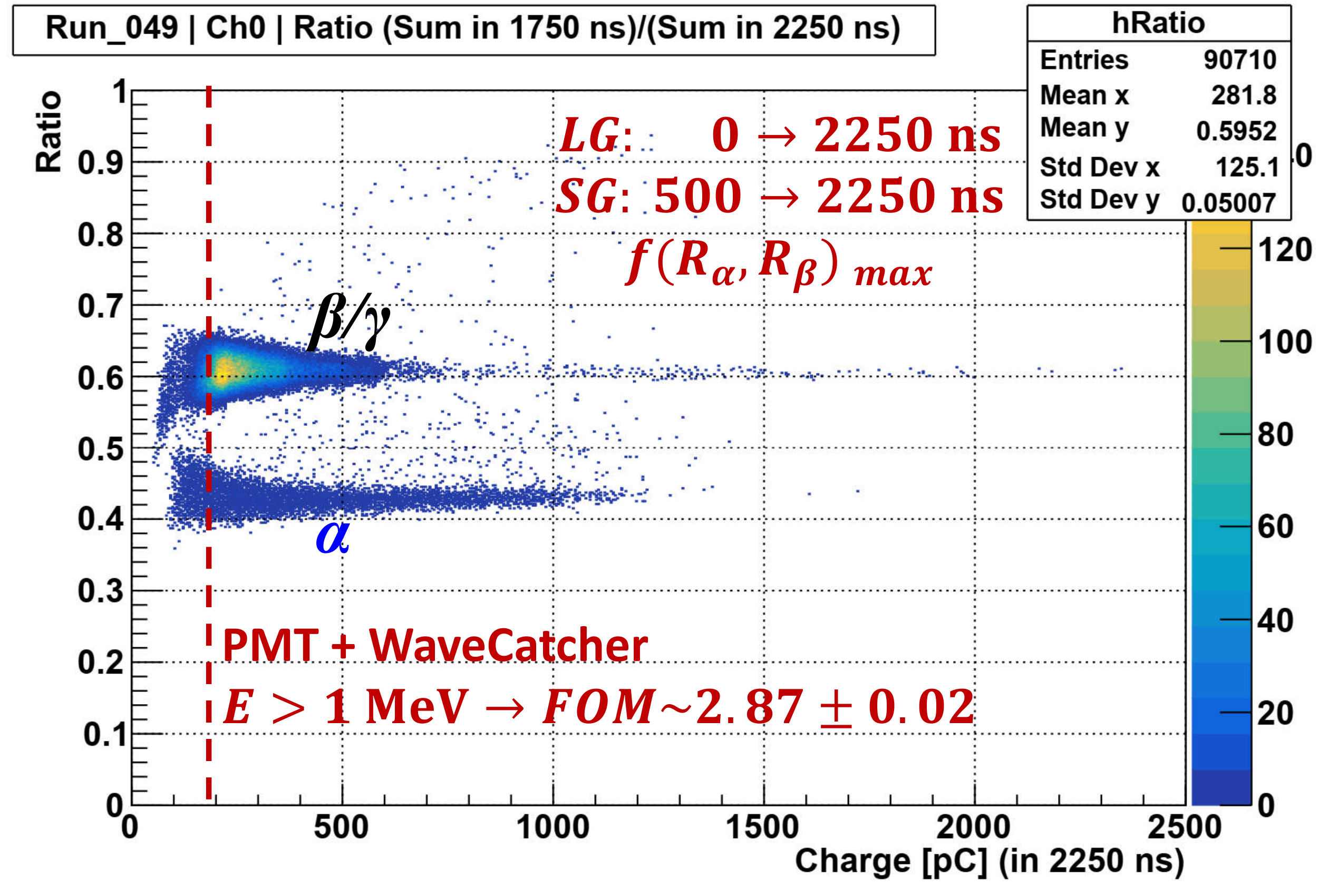}
\includegraphics[width=0.9\linewidth]{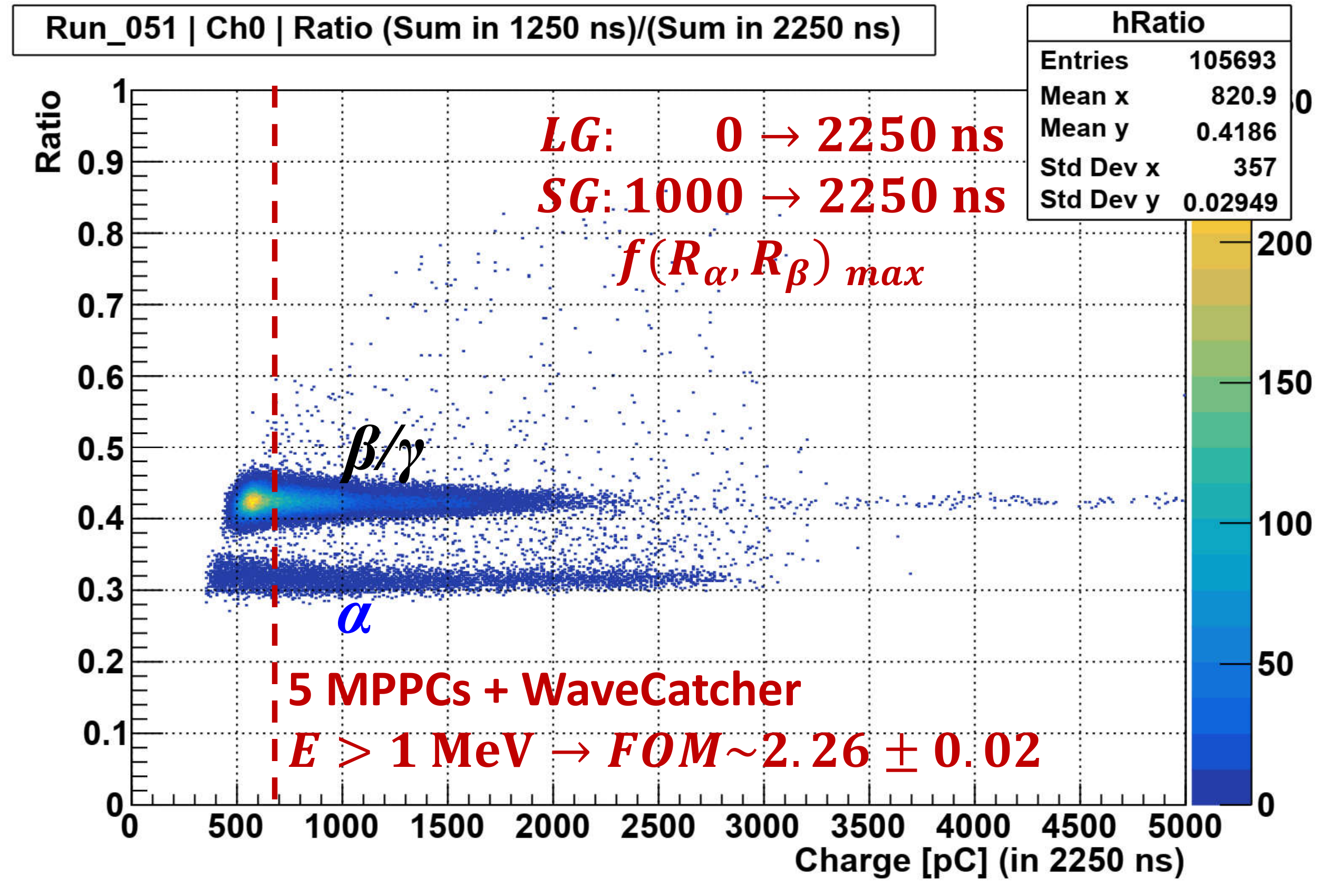}
\caption{The Ratio vs. the charge of the PMT setup (top) and the MPPC setup (bottom) at the estimated SG for $f(R_\alpha,R_\beta)_{max}$.}
\label{fig_Ratio_vs_Q_WC}
\end{figure}

From Eq. \ref{Eq_FOM_2}, to maximize the FOM, we must determine the SG and LG for $f(R_\alpha,R_\beta)_{max}$, which can be estimated from the area-normalized waveforms of $\alpha$ and $\beta$.
For the LG, it is preferable to keep it as long as possible because $\Delta R_{max}$ increases when receiving more contribution from the delayed part of the digitized waveform.
Therefore, we fix the LG from 0 to \SI{2250}{\nano\second} (the prior $\sim \SI{300}{\nano\second}$ for pedestal measurements) on WaveCatcher for both the PMT and MPPC setups.
For the SG, we estimate the SG for $f(R_\alpha,R_\beta)_{max}$.
Because the area inside the SG of the area-normalized waveform represents the Ratio R, it is easy to calculate $f(R_\alpha,R_\beta)$ at various SGs (End points of all the SGs fixed at \SI{2250}{\nano\second}).
Fig. \ref{fig_f(Ra,Rb)} shows the $f(R_\alpha,R_\beta)$ vs. $\Delta t$, where $\Delta t$ represents the Start point of the SG.
The $f(R_\alpha,R_\beta)_{max}$ is 0.080 at $\Delta t = \SI{500}{\nano\second}$ for the PMT setup and 0.048 at $\Delta t = \SI{1000}{\nano\second}$ for the MPPC setup.
\begin{figure} 
	\centering
	\includegraphics[width=0.9\linewidth]{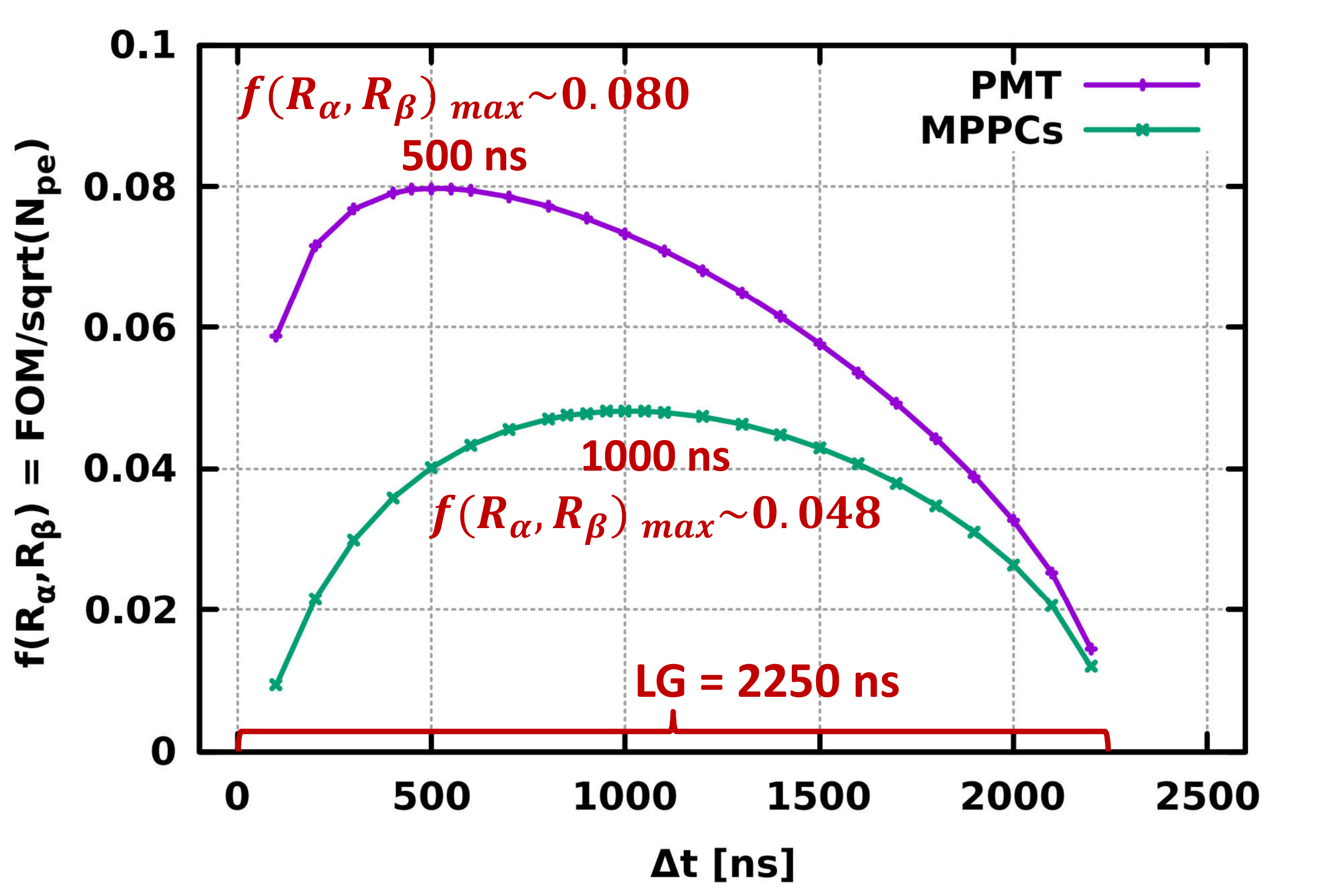}
	\caption{
		Searching the SG for $f(R_\alpha,R_\beta)_{max}$ on both setups: PMT (top) and MPPC (bottom).
		The Start point of the SG represented by $\Delta t = LG - SG$.
		The End point of the SG is fixed at \SI{2250}{\nano\second}.
		The LG is fixed from 0 to \SI{2250}{\nano\second}.
	}	
	\label{fig_f(Ra,Rb)}
\end{figure}
Fig. \ref{fig_Ratio_vs_Q_WC} shows the Ratio vs. the charge at the estimated SG for $f(R_\alpha,R_\beta)_{max}$.
By projecting the events above \SI{1}{\mega\electronvolt} onto the Ratio axis for the Ratio distributions of $\alpha$ and $\beta$, we get the FOM of these two distributions from Gaussian fittings.
The FOM is $2.87 \pm 0.02 $ for the PMT setup and $2.26 \pm 0.02$ for the MPPC setup (above \SI{1}{\mega\electronvolt}).  

%

For the same type of scintillator and the same LG, the $f(R_\alpha,R_\beta)$ curve solely represents the response of the photosensor.
As shown in Fig. \ref{fig_f(Ra,Rb)}, the $f(R_\alpha,R_\beta)$ of the PMT is better than that of the MPPCs in all SGs, $f(R_\alpha,R_\beta)_{max}$ of the PMT $\sim 1.67 f(R_\alpha,R_\beta)_{max}$ of the MPPCs.
This proves that the PMT has a better response than MPPCs.
The faster response time of the PMT is obvious, as demonstrated by the steep rising edges of the two waveforms in Fig. \ref{fig_Nwf_Area_WC} (top).
Therefore, the PMT setup can collect charges more quickly and give better $f(R_\alpha,R_\beta)$.
The slow response of MPPCs compared with that of a PMT was also observed by Grodzicka-Kobylka et al. \cite{grodzicka2018study}, Dinar et al. \cite{dinar2019pulse}, and Budden et al. \cite{budden2012analysis}.
This slow response may result from the capacitance of each pixel inside the MPPC.

\subsection{Optimized SG for $FOM_{max}$} \label{Optimized_SG}
\begin{figure}
	\centering
	\includegraphics[width=0.9\linewidth]{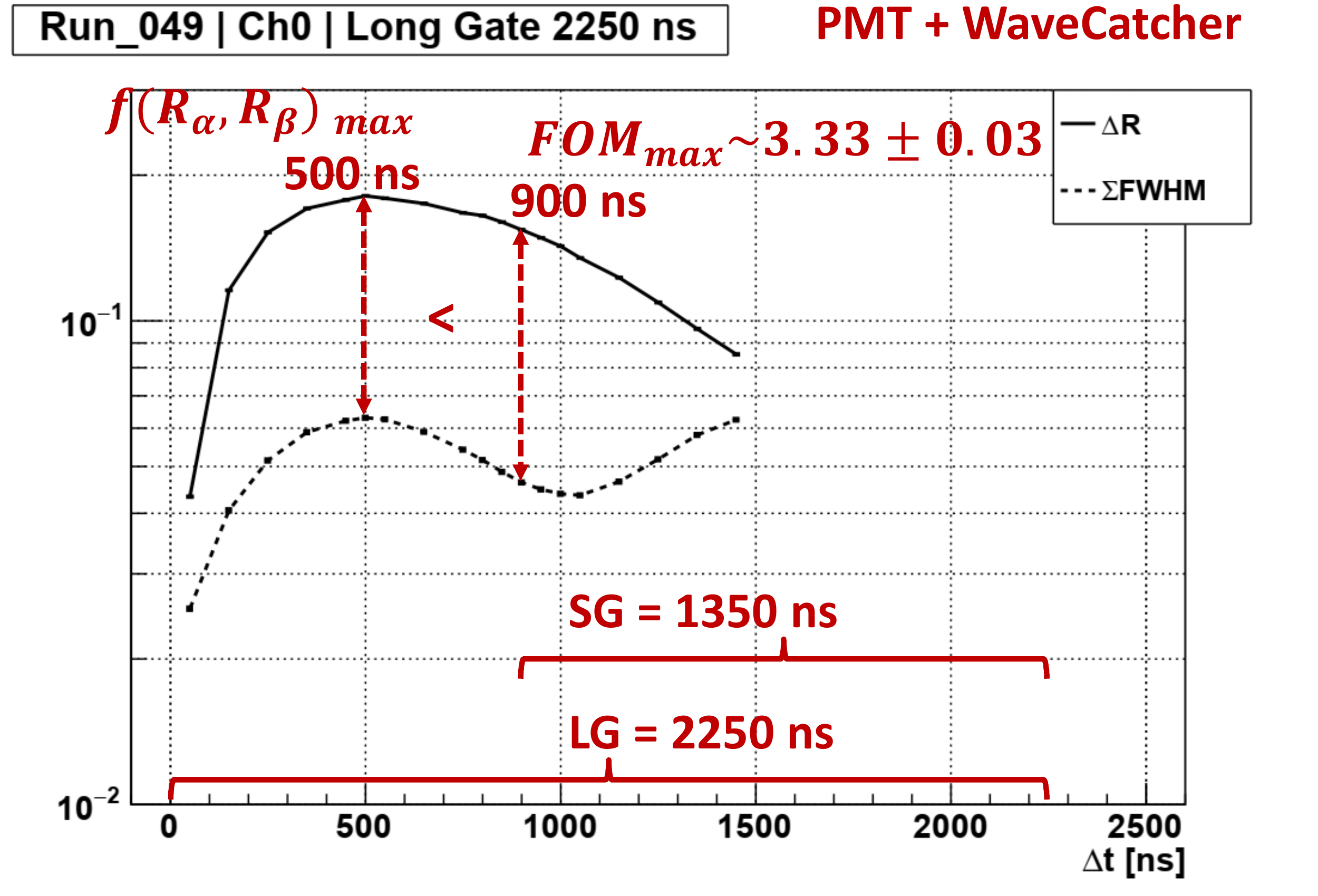}
	\includegraphics[width=0.9\linewidth]{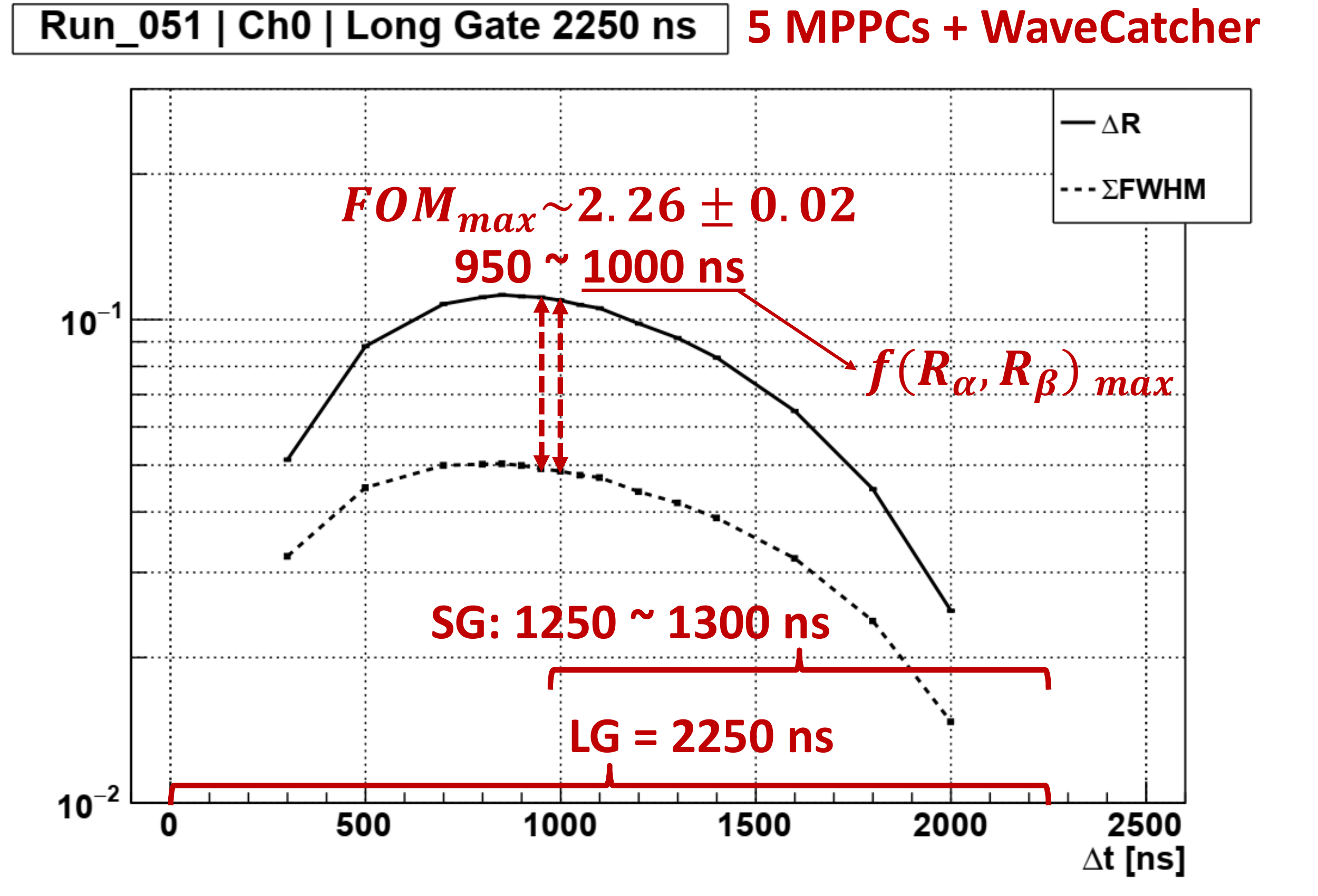}
	\caption{
		Searching the optimized SG for $FOM_{max}$. 
		$\Delta R$ and $\sum FWHM$ (in the FOM definition, Eq. \ref{Eq_FOM_1}) are from fitting. 
		The Start point of the SG is represented by $\Delta t = LG - SG$.
		The End point of the SG is fixed at \SI{2250}{\nano\second}.
		The LG is fixed from 0 to \SI{2250}{\nano\second}.
		In this logarithmic scale, the larger the distance between $\Delta R$ and $\sum FWHM$, the better the FOM.
		For the PMT setup (top), the estimated SG for $f(R_\alpha,R_\beta)_{max}$ is distant from the optimized SG for $FOM_{max}$ as a result of the sinusoidal noise.
		For the MPPC setup (bottom), the estimated SG for $f(R_\alpha,R_\beta)_{max}$ is one of the optimized SGs that give $FOM_{max}$.
	}
	\label{fig_DeltaR_SumFWHM}
\end{figure}

The estimated SG for $f(R_\alpha,R_\beta)_{max}$ in the previous section is the ideal case where the Ratio fluctuation results from the statistical fluctuation of the $N_{pe}$.
To verify whether that SG is optimized for our data, we search the Start point of the SG (Fig. \ref{fig_DeltaR_SumFWHM}) that maximizes the FOM obtained from Gaussian fittings of the two Ratio distributions in our analysis.
Similar to the previous section, the End point of the SG is fixed at \SI{2250}{\nano\second} and the LG is fixed from 0 to \SI{2250}{\nano\second}.

For the PMT setup, the $\sum FWHM$ vs. $\Delta t$ curve indicates the effect of the sinusoidal noise (Fig. \ref{fig_DeltaR_SumFWHM} - top). 
The estimated SG for $f(R_\alpha,R_\beta)_{max}$, which is 500 to \SI{2250}{\nano\second}, is quite distant from the optimized SG for $FOM_{max}$, which is 900 to \SI{2250}{\nano\second}. 
We hypothesize that this optimized SG with the length of \SI{1350}{\nano\second}, which is approximately the period of the sinusoidal noise, mitigates the effect of the noise. 
From fitting, $FOM_{max}$ is $\sim 3.33 \pm 0.03$ at the optimized SG, a noticeable increase from $FOM \sim 2.87 \pm 0.02$ at the estimated SG for $f(R_\alpha,R_\beta)_{max}$.

For the MPPC setup, the estimated SG for $f(R_\alpha,R_\beta)_{max}$, 1000 to \SI{2250}{\nano\second}, and another SG, 950 to \SI{2250}{\nano\second}, both give the $FOM_{max} \sim 2.26 \pm  0.02$ (Fig. \ref{fig_DeltaR_SumFWHM} - bottom). 
Therefore, the SG for $f(R_\alpha,R_\beta)_{max}$ estimated from two area-normalized waveforms gives a satisfactory FOM without an extensive search for the optimized SG in the analysis.
We expect that the PMT setup without the sinusoidal noise will also give a similar result, a satisfactory FOM from the estimated SG for $f(R_\alpha,R_\beta)_{max}$.

\subsection{Charge (or $N_{pe}$) dependence of FOM} \label{FOMvsQ}
\begin{figure}
	\centering
	\includegraphics[width=0.9\linewidth]{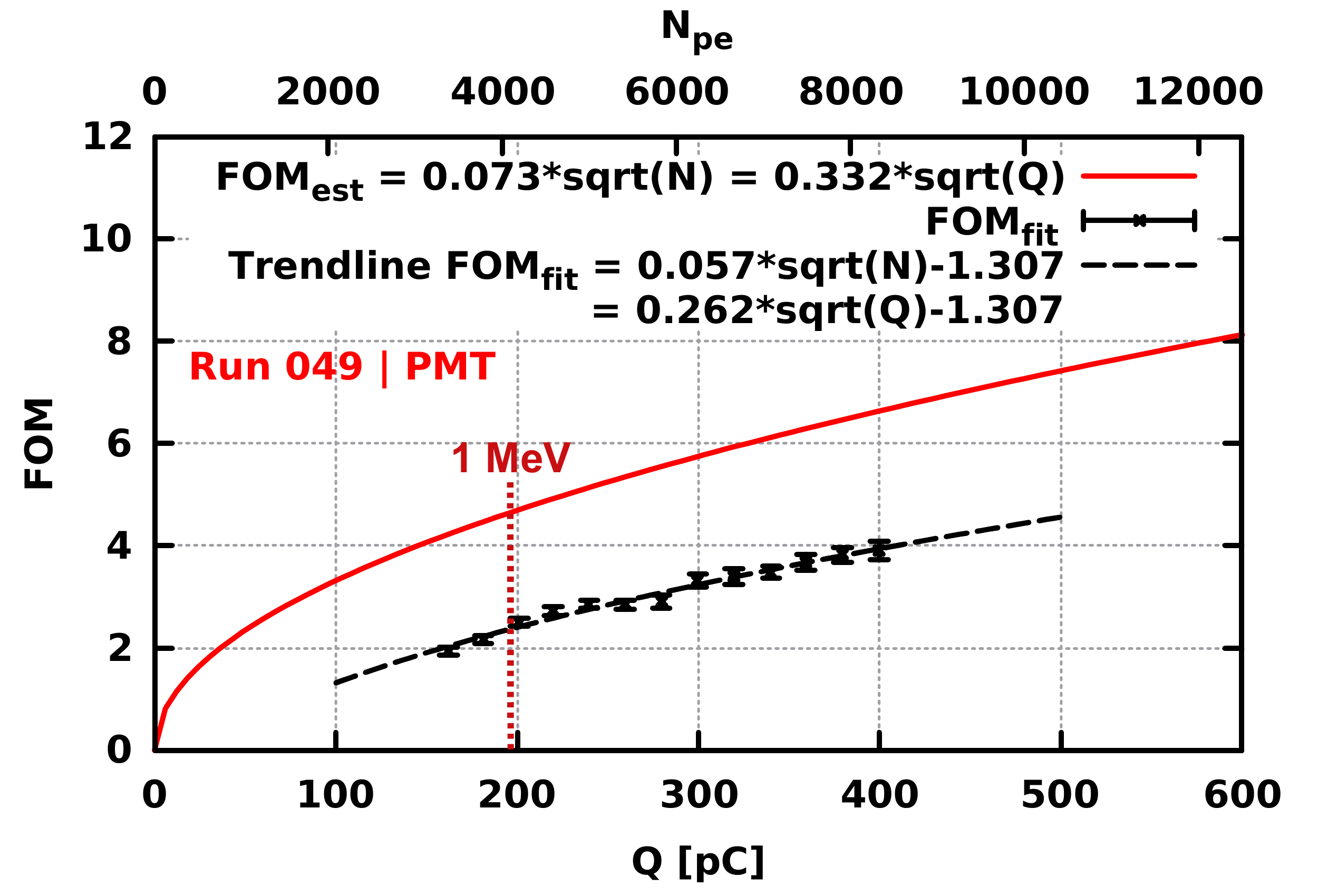}
	\includegraphics[width=0.9\linewidth]{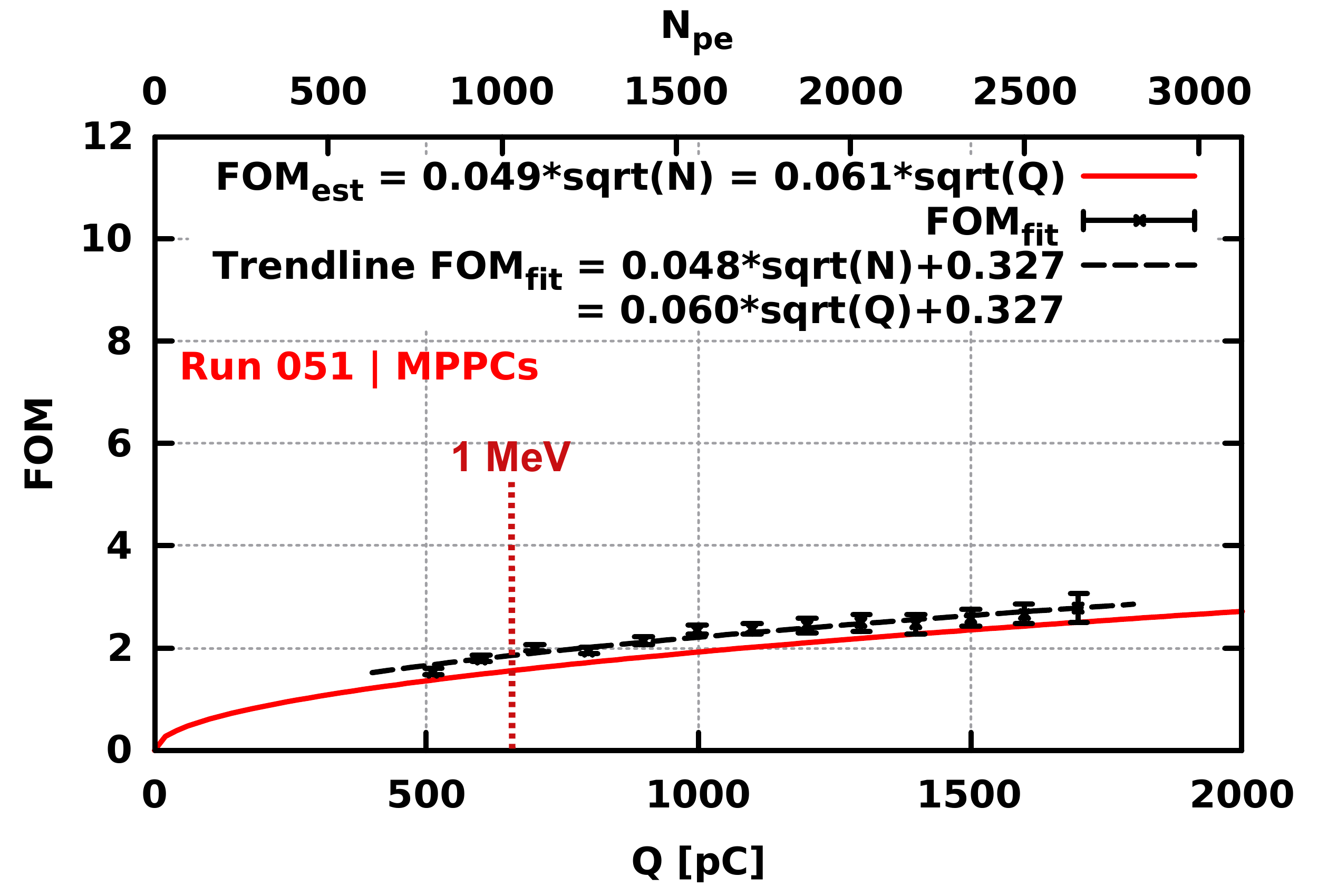}
	\caption{
		The estimated FOM ($FOM_{est}$) vs. $N_{pe}$ and the fitting FOM ($FOM_{fit}$) vs. Q of the PMT setup (top) and the MPPC setup (bottom).
		For the PMT setup, $FOM_{fit}$ is far below $FOM_{est}$ as a result of the sinusoidal noise. 
		For the MPPC setup, the $FOM_{fit}$ is a bit higher than $FOM_{est}$. 
		Despite having noise, the PMT setup still has a better FOM because of a better time response and more $N_{pe}$ generated by the small CsI(Tl) + PMT coupling (about four times the $N_{pe}$ generated by the large CsI(Tl) + five MPPCs coupling; see the figure at \SI{1}{\mega\electronvolt}).
	}
	\label{fig_FOMvsQ}
\end{figure}

In this section, we confirm the relation $FOM \propto \sqrt{N_{pe}}$ or $FOM \propto \sqrt{Q}$ mentioned in section \ref{Est_FOM}.
First, we plot the estimated FOM ($FOM_{est}$) vs. $N_{pe}$ in Eq. \ref{Eq_FOM_2}. 
The values $R_{\alpha}$ and $R_{\beta}$ are from the Ratio distributions at the optimized SG.
The $N_{pe}$ is calculated from the datasheet value of $Q_{1pe}$.
Second, we plot the fitting FOM ($FOM_{fit}$) vs. $Q$ from our analysis result.
Fig. \ref{fig_FOMvsQ} shows the $FOM_{est}$ curve and the discrete values of $FOM_{fit}$ at the optimized SG.

$FOM_{fit}$ tends to be proportional to $\sqrt{Q}$, but it does not match the $FOM_{est}$ curve for both setups.
For the PMT setup, $FOM_{fit}$ is far below $FOM_{est}$.
One reason for this difference is the prior mentioned sinusoidal noise. 
For the MPPC setup, $FOM_{fit}$ is closer to but a bit higher than $FOM_{est}$.

As demonstrated in Eq. \ref{Eq_FOM_2}, to have a better FOM, a setup should have a better time response (greater $f(R_{\alpha},R_{\beta})$) or collect more p.e. (greater $\sqrt(N)$).
From the area-normalized waveform, the PMT proves to have better $f(R_{\alpha},R_{\beta})$.
As shown in Fig. \ref{fig_FOMvsQ}, the small CsI(Tl) + PMT coupling generates about four times the $N_{pe}$ that the large CsI(Tl) + five MPPCs coupling does at the same energy (see Fig. \ref{fig_FOMvsQ} at \SI{1}{\mega\electronvolt}).
Therefore, the PMT setup is better in both time response and p.e. collection; and hence gives a better FOM even being degraded by noise. 
Ideally (without noise), at the same energy, the FOM of the PMT setup is approximately 3.34 (a factor of $0.08/0.048 \sim 1.67$ from $f(R_{\alpha},R_{\beta})_{max}$ and a factor of $\sqrt{4/1}=2$ from $\sqrt{N}$) the FOM of the MPPC setup.
In Fig. \ref{fig_FOMvsQ}, the quotient between the two $FOM_{est}$ is smaller because the $f(R_{\alpha},R_{\beta})$ of the PMT setup, which is taken from the optimized SG, is not maximum.

\subsection{$p-d$ separation} \label{p-d}
As mentioned in Section \ref{Method_Setup}, we plan to use the MPPC setup for the $\mu$ capture on $^3$He experiment in which we will discriminate $p$ and $d$.
Therefore, we perform a $p-d$ separation measurement to evaluate whether the current PSD performance of the MPPC setup is good enough for the $\mu$ capture experiment. 
In this $p-d$ separation, we use a \SI{65}{\mega\electronvolt} proton beam from the AVF (Azimuthal Varying Field) cyclotron at the Research Center for Nuclear Physics (RCNP), Osaka University.
We measure $p$ and $d$ from the reactions of protons with a target of polyethylene and deuterated polyethylene.
We evaluate the result at the scattering angle, $\theta = \SI{45}{\degree}$.

\begin{figure}
	\centering
	\includegraphics[width=0.9\linewidth]{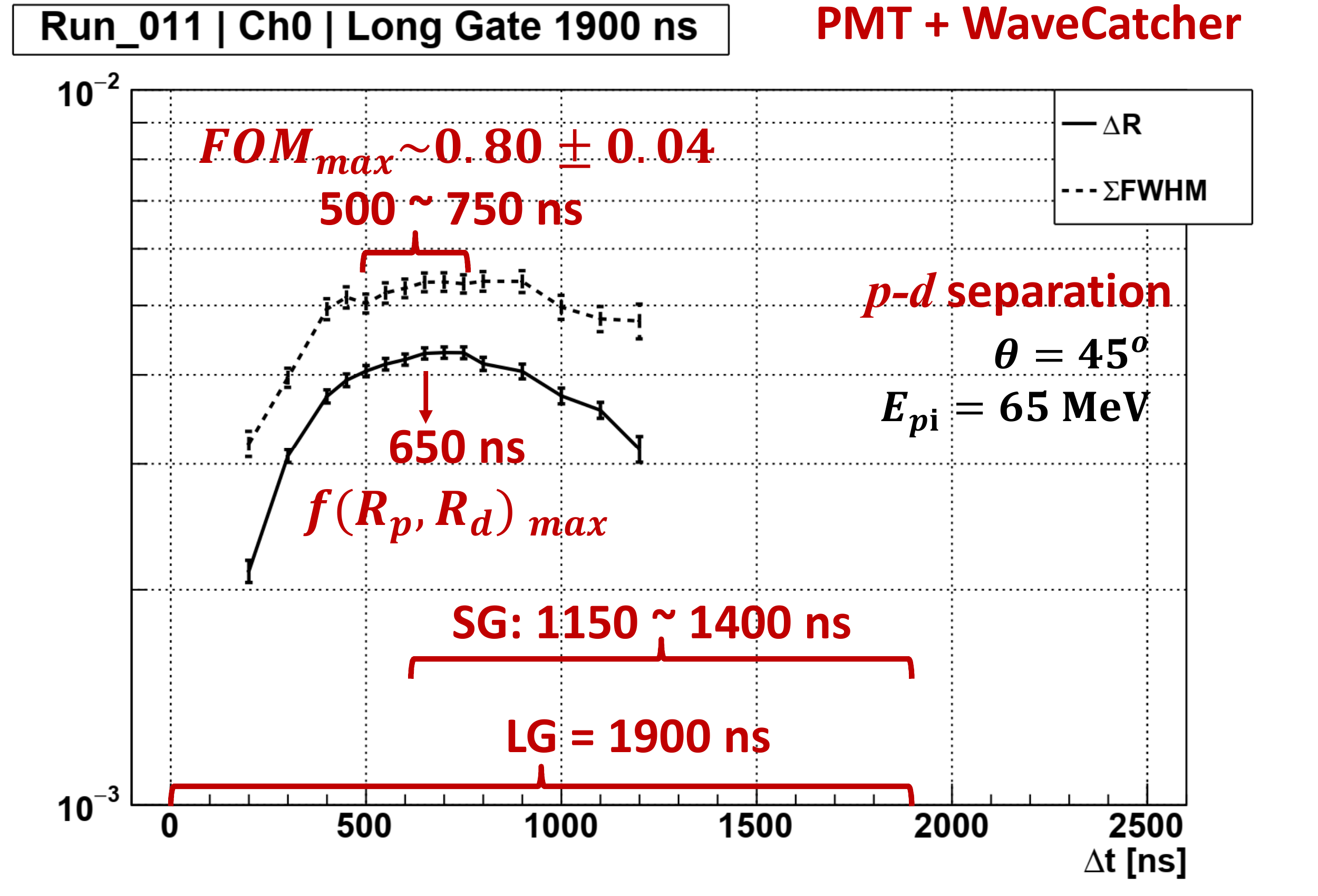}
	\caption{
		Searching the optimized SG for $FOM_{max}$ in the $p-d$ separation using the PMT setup. 
		In this figure, the $\Delta R$ curve is below the $\sum FWHM$ curve.
		Therefore, the closer the distance between $\Delta R$ and $\sum FWHM$, the better the FOM.
		The SG for $f(R_p,R_d)_{max}$ is also one of the SGs giving the $FOM^{p-d}_{max} \sim 0.80 \pm 0.04$.
	}
	\label{fig_DeltaR_SumFWHM_p-d}
\end{figure}

\begin{figure}
	\centering
	\includegraphics[width=80mm]{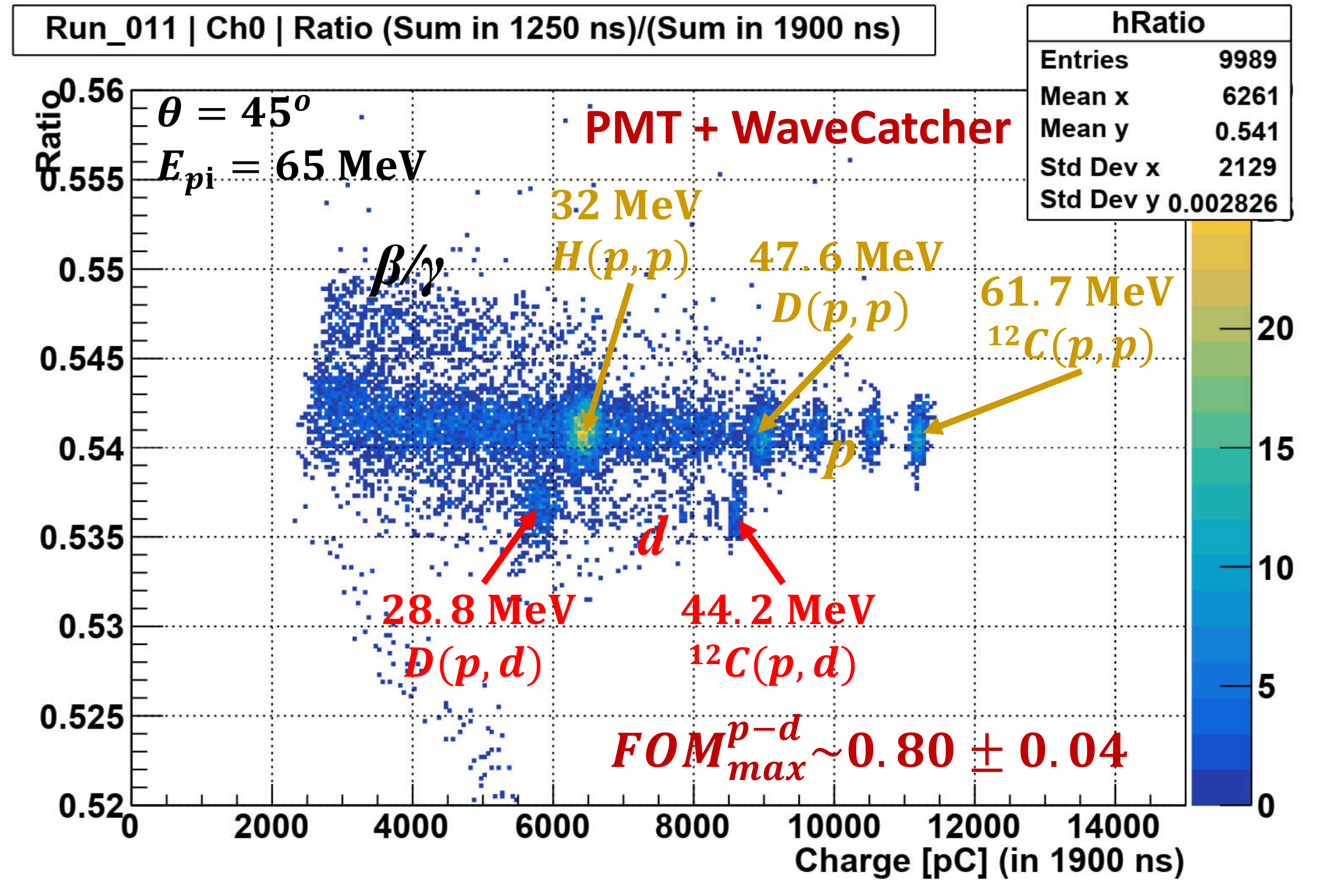}
	\caption{
		The Ratio vs. charge of the $p-d$ separation using the PMT setup at the scattering angle of \SI{45}{\degree}.
		The result is from the SG for $f(R_p,R_d)_{max}$.}
	\label{fig_Ratio_p_d}
\end{figure}

For the PMT setup, we use some modules to adjust the pulse height before feeding the signal into WaveCatcher. 
Unfortunately, we get a refection at the end of the waveform. Therefore, we reduce the LG from 2250 to \SI{1900}{\nano\second}.
We perform a similar analysis, whereby we find the estimated SG for $f(R_p,R_d)_{max}$ from area-normalized waveforms, after which we search the optimized SG for $FOM^{p-d}_{max}$ (Fig. \ref{fig_DeltaR_SumFWHM_p-d}).
The estimated SG for $f(R_p,R_d)_{max}$ is from 650 to \SI{1900}{\nano\second}.
It is one of the optimized SGs (Start points varying at $500 \sim \SI{750}{\nano\second}$ and the End point at \SI{1900}{\nano\second}) that give $FOM^{p-d}_{max} \sim 0.80 \pm 0.04$. 
This good agreement is due to the mitigation of the noise effect at the high energy of $p$ and $d$ (noise is not noticeable in Fig. \ref{fig_DeltaR_SumFWHM_p-d}).
As expected for the PMT setup, when the noise is not significant, the SG for $f(R_p,R_d)_{max}$ can also give a satisfactory FOM without an extensive search for the optimized SG in the analysis. 
Fig. \ref{fig_Ratio_p_d} shows the Ratio vs. charge at the optimized SG.
The $p-d$ separation is quite visible with specific energy peaks.
For the MPPC setup, besides the PSD performance, our experimental condition is also unsatisfactory. Therefore, we cannot separate $p$ and $d$.


\subsection{Improving the PSD performance of MPPC setup by increasing LG} \label{uTCA}
\begin{figure}
	\centering
	\includegraphics[width=0.9\linewidth]{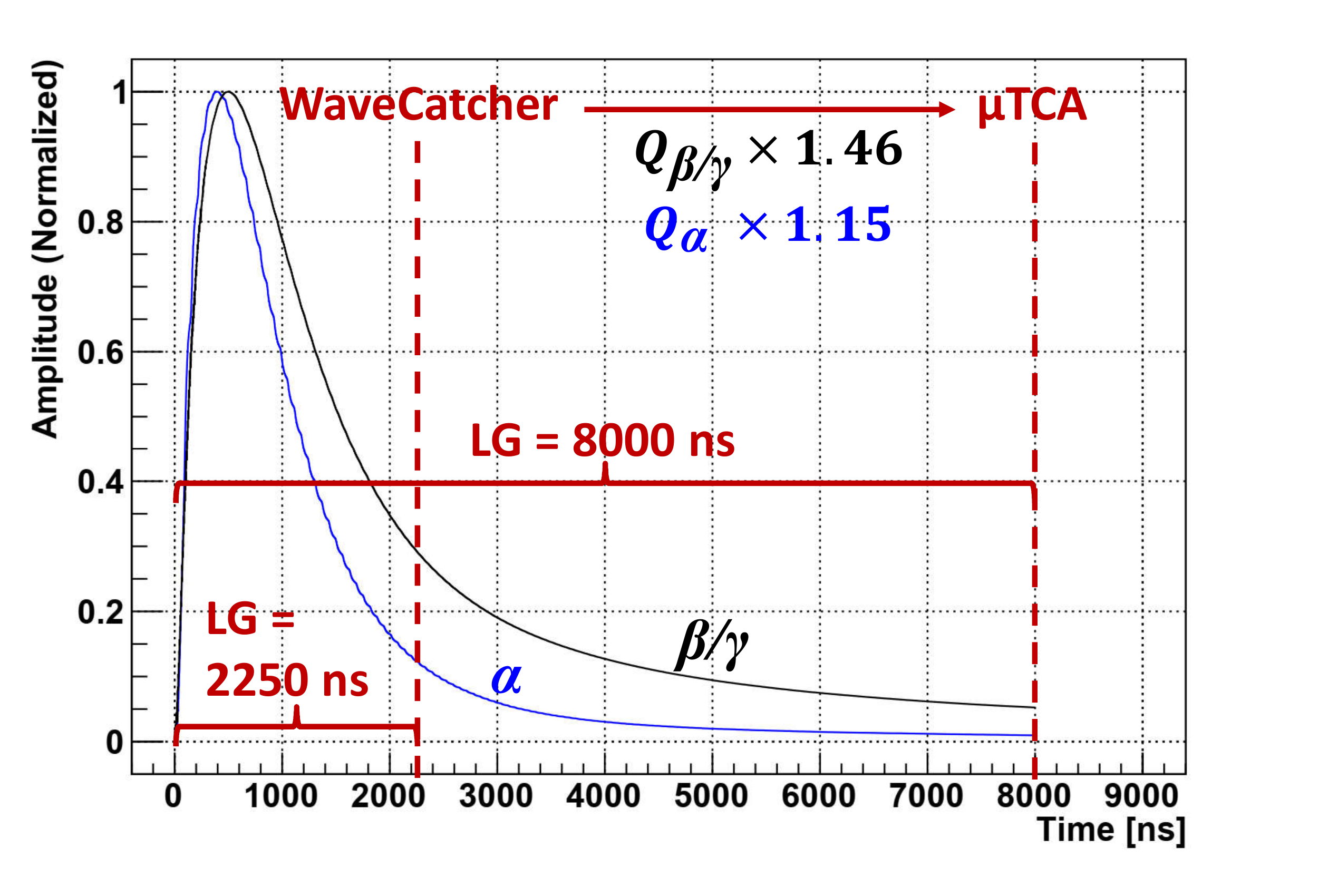}
	\caption
	{
		Amplitude-normalized waveforms of $\alpha$ and $\beta$ from the five MPPCs when using a new DAQ system, the $\mu$TCA.
		The increase in charge collection in the $\mu$TCA system will improve the FOM.
	}
	\label{fig_Nwf_Amp_uTCA}
\end{figure}

Previous sections show that the PSD performance of the MPPC setup is inferior to that of the PMT setup as a result of the slow response time of the MPPC and the low p.e. generation of the large CsI(Tl) + five MPPC coupling.
With $LG = \SI{2250}{\nano\second}$, WaveCatcher can only collect 65\% and 84\% of the total charges of the $\beta/\gamma$ and $\alpha$ events, respectively.
Therefore, we extend the LG for increased charge collection.
This approach will improve the p.e. collection at the delayed part of the digitized waveform (greater $\sqrt{N_{pe}}$) and mitigate the effect of the MPPC's slow response (greater $f(R_\alpha,R_\beta)$).
Eventually, we improve the PSD performance of the MPPC setup.  
To extend the LG to over \SI{2250}{\nano\second} on WaveCatcher, we change to a new data acquisition (DAQ) system using \SI{500}{\mega\hertz} FADC in $\mu$TCA \cite{khai2019mutca}. 
The $\mu$TCA system offers a sampling rate of \SI{0.5}{GS\per\second} with a maximum length for digitized waveforms of \SI{8960}{\nano\second}.
Therefore, we fix $LG = \SI{8000}{\nano\second}$.
With this LG, the $\mu$TCA system can collect up to 95\% and 98\% of the total charges of the $\beta/\gamma$ and $\alpha$ events, respectively (Fig. \ref{fig_Nwf_Amp_uTCA}).

\begin{figure}
	\centering
	\includegraphics[width=0.9\linewidth]{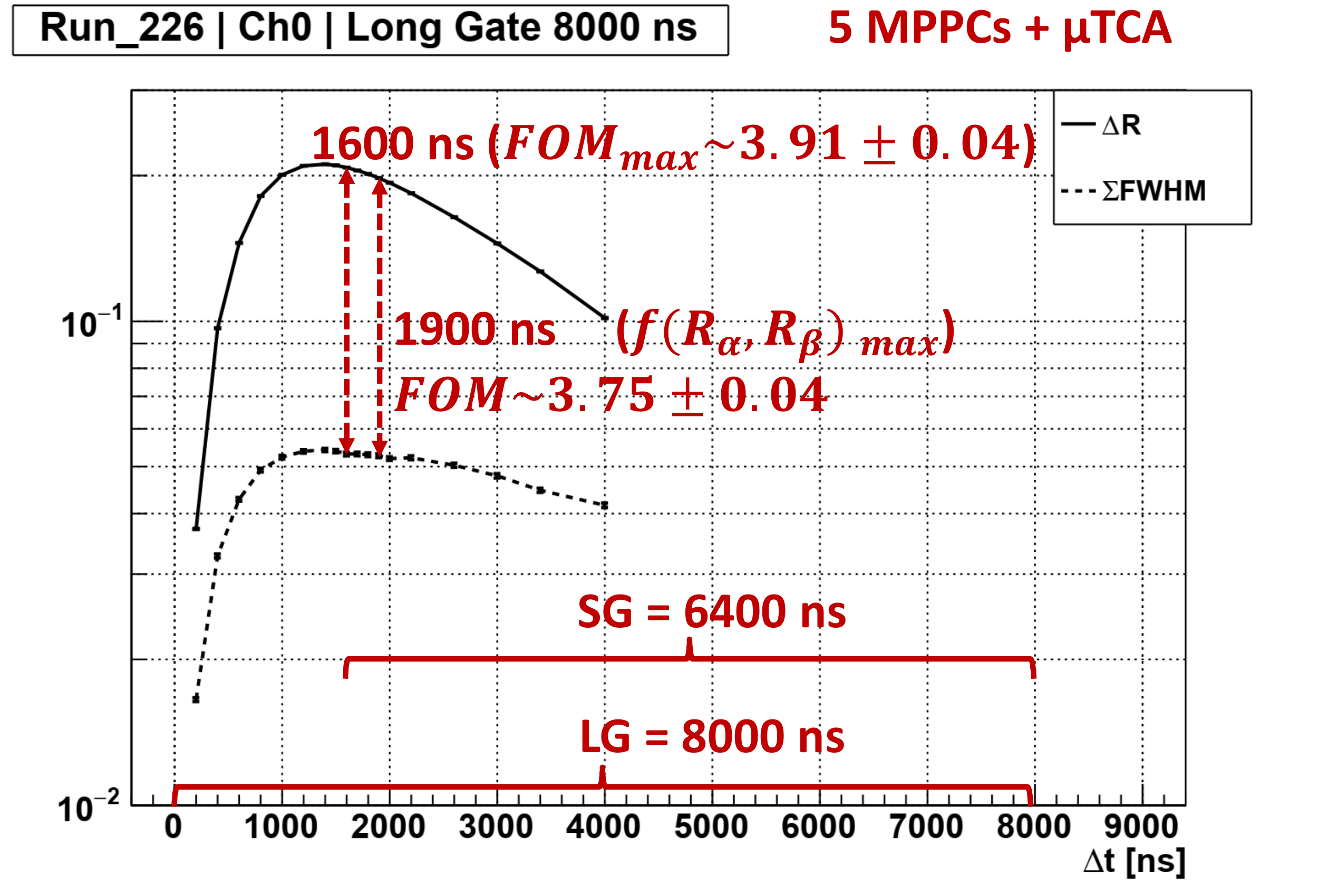}
	\caption{
		Searching the optimized SG for $FOM_{max}$ when using the 5 MPPCs + µTCA.
		The larger the distance between $\Delta R$ and $\sum FWHM$, the better the FOM.
		The estimated SG for $f(R_\alpha,R_\beta)_{max}$ still gives a satisfactory $FOM \sim 3.75 \pm 0.04$, compared to the $FOM_{max} \sim 3.91 \pm 0.04$ at the optimized SG.
	}
	\label{fig_DeltaR_SumFWHM_uTCA}
\end{figure}

\begin{figure}
	\centering
	\includegraphics[width=0.9\linewidth]{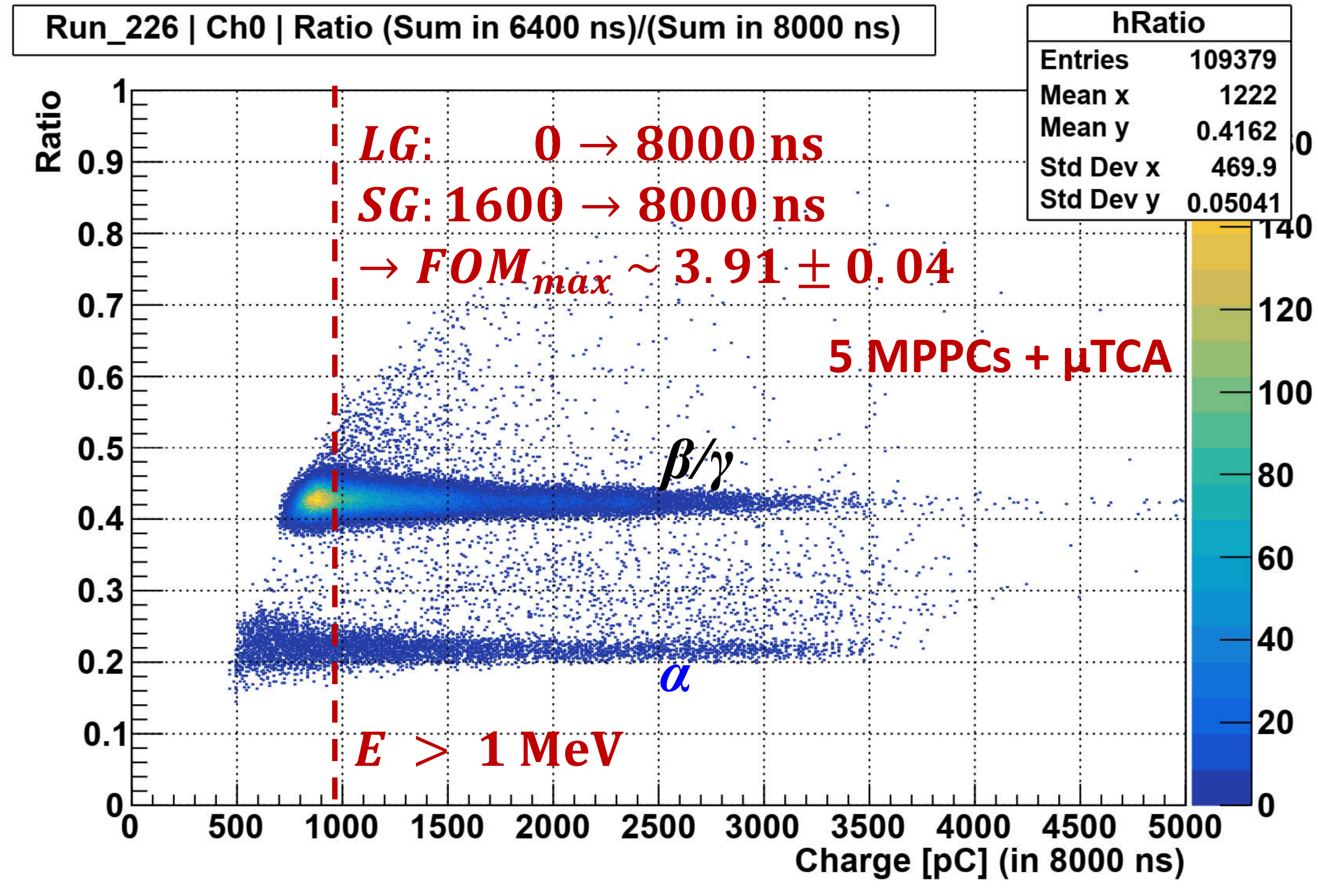}
	\caption{
		The Ratio vs. charge of the $\alpha-\beta$ separation  when using 5 MPPCs + $\mu$TCA. 
		At the optimized SG, the value of $FOM_{max}$ is $\sim 3.91 \pm 0.04$.}
	\label{fig_Ratio_uTCA}
\end{figure}

We perform a similar $\alpha-\beta$ separation measurement using the small CsI(Tl) + five MPPCs coupling with the µTCA system. 
Fig. \ref{fig_DeltaR_SumFWHM_uTCA} shows the search for the Start point of the optimized SG (End point fixed at \SI{8000}{\nano\second}) on the $\mu$TCA system.
From the area-normalized waveforms, we estimate that the SG starting from \SI{1900}{\nano\second} gives $f(R_\alpha,R_\beta)_{max} \sim 0.097$.
This estimated SG is not too far from the optimized SG for $FOM_{max}$ starting from \SI{1600}{\nano\second} that we get from the search.
From fitting, at the estimated SG for $f(R_\alpha,R_\beta)_{max}$, the FOM is $\sim 3.75 \pm 0.04$, which is still satisfactory compared to the $FOM_{max} \sim 3.91 \pm 0.04$ at the optimized SG (the range is above \SI{1}{\mega\electronvolt}). 
These satisfactory FOMs prove that the PSD performance of the MPPC setup improves significantly when getting more p.e. through longer charge integration.
Fig. \ref{fig_Ratio_uTCA} shows the Ratio vs. charge at the optimized SG.

\begin{figure}
	\centering
	\includegraphics[width=0.9\linewidth]{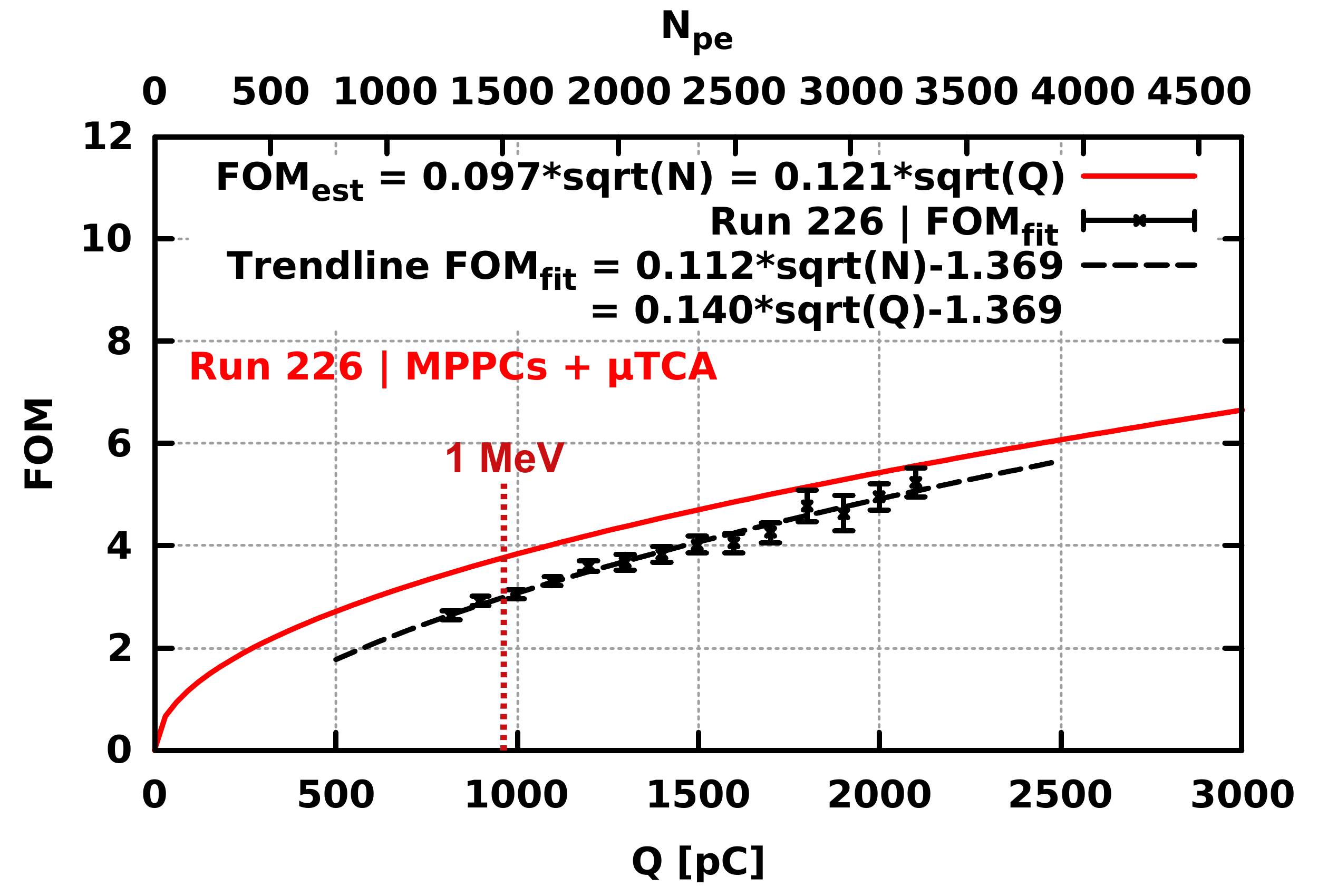}
	\caption{
		The estimated FOM ($FOM_{est}$) vs. $N_{pe}$ and the fitting FOM ($FOM_{fit}$) vs. Q when using five MPPCs + $\mu$TCA.
		The $FOM_{fit}$ values are close to the $FOM_{est}$ curve. 
		Although the FOM of the five MPPCs + $\mu$TCA setup increases significantly, it is still lower than the $FOM_{est}$ (without noise) of the PMT + WaveCatcher setup (in Fig. \ref{fig_FOMvsQ} - top) at the same energy range.
		However, at the same number of p.e., the five MPPCs + $\mu$TCA setup has a better FOM.
	}
	\label{fig_FOMvsQ_uTCA}
\end{figure}

We also confirm the $FOM \propto \sqrt{Q}$ relation in Fig. \ref{fig_FOMvsQ_uTCA} (values are from the optimized SG for $FOM_{max}$).
The $FOM_{fit}$ values are slightly below the $FOM_{est}$ curve.
The $f(R_\alpha,R_\beta)_{max} \sim 0.097$ of the MPPCs + $\mu$TCA setup (also achieved by the optimized SG in Fig. \ref{fig_FOMvsQ_uTCA}) is double the value from the MPPCs + WaveCatcher setup (0.048) and also greater than the value from the PMT + WaveCatcher setup (0.08). 
Ideally, at the same $N_{pe}$, the MPPCs + $\mu$TCA setup will give a better PSD performance (by a factor of $0.098/0.08 = 1.225$) than the PMT + WaveCatcher setup without noise.

Although the $\mu$TCA system can collect more p.e. ($\sim 1.5$ the $N_{pe}$ collected by WaveCatcher), the total $N_{pe}$ generated by the large CsI(Tl) + five MPPCs coupling does not change.
The $N_{pe}$ collected by the MPPCs + $\mu$TCA setup is still inferior to that of the PMT + WaveCatcher setup (by a factor of $4/1.5 \sim 2.67$).   
Therefore, at the same energy range, the PMT + WaveCatcher setup without noise will give a better PSD performance (by a factor of $0.08/0.098 \times \sqrt{4/1.5} \sim 1.33$) than the MPPCs + $\mu$TCA setup.
This indicates that we must improve the large CsI(Tl) + MPPCs coupling for increased scintillation-photon collection.

\section{Conclusion} \label{conclusion}
For a satisfactory PSD performance, an essential factor is the charge collection. 
The long time-gate (covering the entire digitized waveform) for charge integration should be long enough to allow for the collection of almost all the charges generated by the photosensor.
This will result in increased p.e. collection and will effectively improve the response of the photosensor (in the case of the MPPC's slow response) for a satisfactory PSD performance. 

Moreover, in our study, using the area-normalized waveform, we can estimate the short time-gate (delayed part of the digitized waveform) for a satisfactory FOM without an extensive search of this time-gate in the analysis.

\bibliographystyle{ieeetr}
\bibliography{Manuscript_TNS_v6Arxiv}

\end{document}